\documentclass[11pt]{article}

\usepackage[utf8]{inputenc}
\usepackage[T1]{fontenc}

\usepackage[all,2cell,matrix,curve,arrow,tips,frame]{xy}
\usepackage{epsfig}
\usepackage[utf8]{inputenc}
\usepackage[T1]{fontenc}
\usepackage{amsmath}
\usepackage{amsfonts}
\usepackage{amsthm}
\usepackage{epic}
\usepackage{eepic}
\usepackage{amssymb}
\usepackage{mathrsfs}
\usepackage{graphicx}
\usepackage{amsmath}
\usepackage{algorithmic}
\usepackage{algorithm}

\newcommand{\edge}[1]{\ar@{-}[#1]}

\def\thebibliography#1{\section*{Literature Cited\markboth
{Literature Cited}{Literature Cited}}\list
{[\arabic{enumi}]}{\settowidth\labelwidth{[#1]}\leftmargin\labelwidth
\advance\leftmargin\labelsep 
\usecounter{enumi}}
\def\newblock{\hskip .11em plus .33em minus -.07em}
\sloppy
\sfcode`\.=1000\relax}

\newtheorem{Th}{Theorem}

\newtheorem{Lmm}[Th]{Lemma} 
 
\newtheorem{Defn}[Th]{Definition}

\begin{document}

\begin{center}

{\LARGE 
An Algorithm for Learning the Essential Graph}\vspace{5mm}

John M. Noble\\ 
Matematiska institutionen,\\
Linköpings universitet,\\
58183 LINKÖPING, Sweden\\
email: jonob@mai.liu.se
\end{center}

\noindent Condensed running title: Modified Maximum Minimum Parents and Children Algorithm

\begin{abstract} This article presents an algorithm for learning the essential graph of a Bayesian network. The basis of the algorithm is the Maximum Minimum Parents and Children algorithm from~\cite{TBA}, with three substantial modifications. The MMPC algorithm is the first stage of the Maximum Minimum Hill Climbing algorithm  for learning the directed acyclic graph of a Bayesian network, introduced in~\cite{TBA}. The MMHC algorithm runs in two phases; firstly, the MMPC algorithm to locate the skeleton and secondly an edge orientation phase. The computationally expensive part is the edge orientation phase. 

The first modification introduced to the MMPC algorithm, which requires little additional computational cost, is to obtain the immoralities and hence the essential graph. This renders the edge orientation phase, the computationally expensive part, unnecessary, since the entire Markov structure that can be derived from data is present in the essential graph. 

Secondly, the MMPC algorithm can accept independence statements that are logically inconsisted with those rejected, since with tests for independence, a `do not reject' conclusion for a particular independence statement is taken as `accept' independence. An example is given to illustrate this and a modification is suggested to ensure that the conditional independence statements are logically consistent.

Thirdly, the MMHC algorithm makes an assumption of faithfulness. An example of a data set is given that does not satisfy this assumption and a modification is suggested to deal with some situations where the assumption is not satisfied. The example in question also illustrates problems with the `faithfulness' assumption that cannot be tackled by this modification.
\end{abstract}

\section{Introduction}\label{secintro} To ensure that the paper is self contained, the definitions and main results used, that have been developed by other authors, will be stated in the appendix (section~\ref{app}). This article considers probability distributions over a set of random variables $V = \{X_1, \ldots, X_d\}$ that may be represented by a Bayesian network (definition~\ref{defBN}). Each variable is multinomial and the variables have a dependency structure that is unknown and is to be established from data. The algorithm in this article is intended for use for networks used to model domain knowledge in Decision Support Systems, particularly in medicine. An introduction to the topic is found in~\cite{KN}, a more advanced treatment in Cowell et. al.~\cite{CDLS} and applications within medicine in ~\cite{BSCC} and implementation in~\cite{AJAFKW}.

Learning a Bayesian network (that is, locating the graph structure) from data is an important problem in several fields. It has been studied extensively over the last 15 years. A successful structure learning algorithm enables a Decision Support System to be constructed automatically. In bioinformatics,  Bayesian networks learned from data have been used for locating gene regulatory pathways. This is discussed by Friedman et. al. in~\cite{FLNP}. In some cases, for example with gene pathways, structure learning algorithms are used to infer causality. When causality is being learned, the essential graph (definition~\ref{defesgr}) is more informative than a directed acyclic graph, because all the edges and only those edges for which the direction of an edge in the directed acyclic graph may be inferred from data are directed in the in the essential graph.   Discussion of the use of Bayesian networks for inferring causality is discussed in~\cite{SGS}; clearly causality cannot be inferred purely from data. Such inferences can only be made if there are other assumptions, for example, that the data comes from a controlled experiment where the candidates for cause to effect are already clear. If only the data is used, then only those arrows that are directed in the essential graph have a clear causal interpretation and the causal link may only be drawn if there are additional modelling assumptions.  

Structure learning is also of importance for classification and variable selection problems as discussed by Tsamardinos and Aliferis~\cite{TA} and structure learning techniques provide the basis of algorithms for solving these problems~\cite{TAS1, TAS2}. 

Biomedical sciences and genetics, with micro-array gene expression techniques, often produce data sets with tens or hundreds of thousands of variables, which require algorithms that enable learning of a Bayesian network in real time. Learning Bayesian networks from data, without any additional assumptions (for example, that there exists a directed acyclic graph that is faithful (definition~\ref{deffaith}) to the distribution, or limiting the number of possible nodes in each parent set) is an NP-hard problem (see~\cite{Ch, ChMH}). 

In general, there are two main approaches for learning Bayesian networks from data; search and score techniques and constraint based techniques. A search and score techniques looks for the network that maximises a score function, where the score function indicates how well the network fits the data. The score function may be a likelihood function, or a posterior probability function $p_{{\cal E}| {\underline X}}(E| {\bf x})$ where $E$ is a structure and ${\bf x}$ is the data matrix. These algorithms search the space of possible structures for the structure $E$ that maximised the score function, using a greedy algorithm, local algorithm or some other search algorithm. These are discussed in Cooper and Herskovitz~\cite{CH}, Heckerman, Geiger and Chickering~\cite{HGC}. 

Constraint based algorithms are discussed in Spirtes, Glymour  and Scheines~\cite{SGS}. These algorithms investigate from the data whether or not there are certain conditional independence structures between the variables. This is performed using statistical or information theoretic measures. Networks that do not have all the conditional independence statements ascertained from the algorithms are eliminated from consideration.

The constraint - based approach, found in the MMPC algorithm (the skeleton construction stage) of~\cite{TBA} has a problem that is  illustrated with the example in subsection~\ref{subsWAM} with $1000$ observations on $6$ binary variables. The statement `do not reject independence' is taken as `accept independence' and this acceptance is made without reference to other independence statements that have been rejected. The  example (subsection~\ref{subsWAM})) gives a situation where the  statement $X \perp Y | \{W,Z\}$ is not rejected   and hence $X \perp Y | \{W,Z\}$ is accepted, so that the edge $\langle X, Y \rangle$ is then removed from the graph, even though $X \not \perp Y$, $X \not \perp Y | Z$ and $Y \perp\{W,Z\}$. It is a relatively easy computation to establish that  
\[ Y \perp \{W,Z\}\qquad \mbox{and}\qquad X \perp Y | \{W,Z\} \Rightarrow X \perp Y | Z\] 
\noindent and therefore the statement $X \perp Y | \{W,Z\}$ is incompatible with the statement $X \not \perp Y | Z$ that has already been accepted. 

The problem is that `do not reject' only means `there is insufficient evidence based on this single hypothesis test to reject.' It is therefore wrong to {\em accept} the independence statement if the rejection of other statements logically leads to the rejection of $X \perp Y | \{Z,W\}$. 

The modification suggested here takes a hierarchical approach; a statement $X \perp Y | S$ is accepted if and only if the test statistic is sufficiently low and it is also consistent with all the conditional independence statements that may be derived from the data involving smaller conditioning sets. 

This modification increases the computational complexity. This increase may not be serious in very large examples, with a very large number of variables and a very large number of instantiations, where the time taken for a statistical call (definition~\ref{defstatcall}) is essentially due to computing the cell counts for the subsets of the variables involved, and once the cell counts for these variables have been computed, the time taken to make the computation for subsets of these variables is insignificant. For smaller examples, the increase in complexity may be substantial.

In the small six variable example in subsection~\ref{subsWAM}, the modification gives a substantial improvement to  the quality of the fitted distribution.   

The Maximum Minimum Parents and Children algorithm is introduced in~\cite{TBA}, to find the skeleton (definition~\ref{defskel}) of a directed acyclic graph that describes the independence structure. Following the discussion above, it is constraint based; it removes an edge $\langle X, Y \rangle$ from the skeleton if there is a set of variables $S$ such that $X \perp Y | S$. It works in a similar manner to the PC algorithm, developed by Spirtes, Glymour and Scheines in~\cite{SGS}.

The algorithm requires that   there exists a directed acyclic graph that is {\em faithful} (definition~\ref{deffaith}) to the probability distribution. That is, all conditional independence statements in the probability distribution are represented by $d$-separation statements (definition~\ref{defdsep}) in the graph. The definition  of {\em faithful} is found in~\cite{KN} page 69 while the definition of {\em $d$-separation} is found in~\cite{KN} page 49. These notions were developed and discussed in  work  by J.Pearl, D.Geiger and T. Verma, for example~\cite{PV} and~\cite{PGV}.

In motivating the `faithfulness' assumption in~\cite{TBA}, they comment in section 2 (under the statement of theorem 1 in that article) that there are distributions for which there do not exist  {\em faithful} Bayesian networks, but that these distributions are rare, as discussed by Meek in~\cite{Me}. But the Women and Mathematics example (subsection~\ref{subsWAM}) was a randomly chosen example on 6 variables, chosen only because of its convenience in size for a toy model and because it provided a useful comparison - other authors had already used it to test structure learning algorithms. It turned out that the dependence structure  derived from data, where conditional independence is rejected when the null hypothesis is rejected at the $5\%$ significance level, did not correspond to a distribution that had a faithful graphical representation. 

There are two ways that lack of faithfulness can appear. The first is where an algorithm based on the principle of adding in a edge in the skeleton between $X$ and $Y$ when and only when there is no conditioning set $S$ such that $X \perp Y | S$ produces a skeleton that can represent all the associations, but any DAG will necessarily have $d$ connection statements that do not correspond to conditional dependence. This suggests associations that do not exist, but the procedure developed in this article will produce a representation that gives an accurate model for the underlying distribution, in the sense of Kullback Leibler distance. 

More seriously, a distribution does not satisfy the faithfulness assumption if there is a situation where variable $A$ is independent of $C$ and variable $B$ is independent of $C$, but variables $A$ and $B$ taken together tell us everything about $C$. This (to quote Edward Nelson~\cite{Nelson}) is the principle on which any good detective novel is based and joint distributions that have this property are not rare, at least within the social sciences. 

The example presented in section~\ref{subsWAM} is a randomly chosen example from the social sciences containing a group of variables that have this property.

All  graphs faithful to a given set of conditional independence statements are Markov equivalent (definition~\ref{defmeq}) to each other. All the graphs in a  set of directed acyclic Markov equivalent graphs  have the same skeleton and the same immoralities (theorem~\ref{thmeq}). This result was established by T.Verma and J. Pearl in~\cite{PV3}. These ideas are discussed in~\cite{KN} sections 4.4 and 4.5; the works of D. Geiger, J. Pearl and T.Verma~\cite{PV3} and~\cite{GPV2} are important in this regard, as are important contributions by the authors D. Madigan, S.A. Andersson, M.D. Perlman and C.T. Volinsky (for example,~\cite{MAPV}). Therefore, if there is a directed acyclic graph faithful to the distribution, then all faithful DAGs have the same skeleton and this unique skeleton is returned by the MMPC algorithm. 

The Maximum Minimum Hill Climbing algorithm was introduced in~\cite{TBA}. It consists of two stages; firstly the MMPC algorithm  and secondly, having located the skeleton, a stage where the edges of the skeleton are oriented to find the best fitting DAG  that describes the conditional independence statements of a probability distribution. In general, if a distribution has a faithful representation, there will be several possible faithful DAGs; any graph within the Markov equivalence class determined by the conditional independence statements will fit the data equally well. Only the {\em immoralities} (definition~\ref{defimm}) and {\em strongly protected edges} (definition~\ref{defstpred}) are the same in every faithful DAG. The background material is discussed in~\cite{KN} sections 4.4 and 4.5; see ~\cite{MAPV} and ~\cite{GPV2}. There is therefore a level of  inefficiency in an algorithm that operates by altering one edge at a time, adding a directed edge, or removing a directed edge, or changing the direction of a directed edge, that aims to find a suitable DAG; it may spend time comparing different DAGs that have the same essential graph.   

This article discusses a modification of the Maximum Minimum Parent Child algorithm that   ensures that no conditional independence statements incompatible with known dependencies are accepted, and which also locates the immoralities. It then provides a straightforward routine to locate the essential graph (definition~\ref{defesgr}). 

A substantial  evaluation of MMHC against the most widely used  Bayesian network learning algorithms is given in~\cite{TBA}. These include the PC algorithm (Spirtes, Glymour and Scheines~\cite{SGS}), the Three Phase Dependency Analysis (Cheng, Greiner, Kelly, Bell and Liu~\cite{CGKBL}), the Sparse Candidate (Friedman, Nachman and Pe'er~\cite{FNP}), Optimal Reinsertion (Moore and Wong~\cite{MW}), Greedy Equivalent Search (Chickering~\cite{Ch2}) and Greedy Search.

\section{Background} The background material is given extensively in the appendix, section~\ref{app}, to keep the article self contained; this section is restricted to a few brief remarks and some notations.  The variables, and nodes of a graph are denoted by upper case letters, with indices where appropriate; for example, $X$, $X_i$. Sets of variables are denoted by upper case letters; for example, $S$. The meaning will be clear from the context. The probability function for a collection of random variables $X_1, \ldots, X_d$ will be denoted $p_{X_1, \ldots, X_d}$, a function of $d$ variables. An instantiation of a variable will be denoted by the same letter in lower case, e.g. $x_i$ for an instantiation of variable $X_i$. Hence $p_{X_1,\ldots, X_d}$ is a function of $d$ variables  and $p_{X_1,\ldots, X_d}(x_1,\ldots, x_d)$ denotes the probability that the random vector $(X_1,\ldots, X_d)$ take the value $(x_1,\ldots, x_d)$. 

Two variables $X$ and $Y$ are conditionally independent given a set of variables $S = \{Z_1, \ldots, Z_m\}$ if for each $(x,y,z_1, \ldots, z_m)$
\[p_{X,Y,Z_1,\ldots, Z_m}(x,y,z_1,\ldots, z_m) = p_{X|Z_1,\ldots, Z_m}(x|z_1,\ldots, z_m)p_{Y|Z_1,\ldots, Z_m}(y|z_1, \ldots, z_m)p_{Z_1,\ldots, Z_m}(z_1, \ldots, z_m)\]

\noindent where $p_{X|Z_1,\ldots, Z_m}$ denotes the conditional probability function of $X$ given $(Z_1, \ldots, Z_m)$. The notation $X \perp Y$ denotes $X$ independent of $Y$ and the notation $X \perp Y | S$ denotes $X$ independent of $Y$ conditioned on a set of variables $S$. Similarly, for a single variable $Z$, $X \perp Y| Z$ denotes $X$ independent of $Y$ given $Z$ and for sets of variables $A,B,S$, $A \perp B|S$ denotes the variables in set $A$ are jointly independent of the variables in set $B$ given set $S$. The notation $\not \perp$ denotes the negation of $\perp$; for example, $X \not \perp Y | S$ denotes that $X$ is not conditionally independent of $Y$ given $S$. 

The  definition of a Bayesian network is given in definition~\ref{defBN}; it involves the factorisation of a probability distribution according to a Directed Acyclic Graph (DAG), together with the DAG. A probability distribution $p_{X_1, \ldots, X_d}$ factorises as
\[ p_{X_1, \ldots, X_d} = \prod_{j=1}^d p_{X_j | \Pi_j} \] 
where for each $j \in \{1, \ldots, d\}$, $\Pi_j \subseteq \{X_1, \ldots, X_{j-1}\}$ and the statement no longer holds if any of the $\Pi_1, \ldots, \Pi_d$ are replaced by strict subsets. To the factorisation is associated a directed acyclic graph ${\cal G} = (V, D)$, where $V = \{X_1, \ldots, X_d\}$ and for each $j$, the parents of $X_j$ are the sets $\Pi_j$. All the definitions are given in section~\ref{app}. 

The connections between variables in a DAG are either fork, collider or chain and their definitions are given in definition~\ref{deffcc}. The definition of $d$-separation, given in definition~\ref{defdsep}; the definition of a trail is given in definition~\ref{deftrail}. The definition of blocked trail is given in definition~\ref{defbltr}; a trail is blocked by a set of nodes $S$ if there is a node $W$ in the trail such that either 
\begin{itemize}
 \item $W$ is not a collider and $W \in S$ or
\item $W$ is a collider and neither $W$ nor any of its descendants are in $S$.
\end{itemize}
This was introduced by Pearl in~\cite{P} in 1988. Two nodes are $d$-separated by a set $S$ if every trail between them is blocked by $S$ and two sets of variables are $d$ separated by $S$. A directed acyclic graph is said to be {\em faithful} to a probability distribution $p$ if and only if every conditional independence statement in the distribution is represented by a $d$-separation statement in the DAG. This is given by definition~\ref{deffaith}.  

In every Bayesian network, a $d$-separation statement in the DAG represents a conditional independence statement in the distribution. There is a direct proof of this in theorem 2.2, page 66 of~\cite{KN}; the result was established by Verma and Pearl (1988)~\cite{VP}.

The MMPC algorithm locates an appropriate skeleton if and only if there exists a DAG ${\cal G}$ such that $p$ and ${\cal G}$ are faithful to each other. Theorem~\ref{thfaithedge} in the appendix (section~\ref{app}), stating that a graph faithful to a probability distribution contains an edge between two nodes $X$ and $Y$ if and only if  $X \not \perp Y | S$ was introduced by Spirtes, Glymour and Scheines in 1993~\cite{SGS93} and is the crucial characterisation to determine the edges that should be absent or present in the skeleton of a faithful graph.

Algorithms following the constraint based approach estimate from the data whether or not certain conditional independence statements hold and, if they do, remove the appropriate edges.    The MMPC algorithm in~\cite{TBA} will remove an  edge $\langle X, Y \rangle$   from the skeleton if $H_0$: $X \perp Y | S$ is not rejected even if this contradicts statements of dependence that have already been established.  The modification of the MMPC algorithm presented in this article for determining conditional independence therefore only accepts  a statement $X \perp Y | S$ provided that 
\begin{enumerate}
 \item the result of the individual hypothesis test is `based on this individual hypothesis test, do not reject the hypothesis' and 
\item accepting this statement does not contradict conditional dependence statements with smaller conditioning sets that have already been established.
\end{enumerate}

The  modification of the MMPC presented in this article to locate the immoralities and hence the essential graph, requires some further background. An immorality, defined in definition~\ref{defimm}, is a collider connection $X \rightarrow Y \leftarrow Z$ where $X-Y-Z$ form a vee structure (definition~\ref{defveestr}). The key results, following  the definition of Markov equivalence (definition~\ref{defmeq}) are that all faithful graphs are Markov equivalent and theorem~\ref{thmeq}, that two DAGs are Markov equivalent if and only if they have the same skeleton and the same immoralities. It follows that the immoralities are the same in all faithful graphs. If there is a structure where the skeleton contains edges $\langle X, Y\rangle$ and $\langle Y,Z \rangle$ but no edge $\langle X, Z \rangle$, then firstly, if the DAG is faithful, there is a set $S$ such that $X \perp Z | S$. Secondly, if $(X,Y,Z)$ is an immorality, then $X \not \perp Z | R \cup \{Y\}$ for any subset $R \subseteq V \backslash \{X,Y,Z\}$. In particular, for a distribution that possesses a faithful graphical representation, once the skeleton has been determined and once a set $S$ such that $X \perp Z | S$ has been located,  then $(X,Y,Z)$ is an immorality in a faithful DAG if and only if $X \perp Z |S$. Furthermore, $X \not \perp Z | S \cup \{Y\}$. 

Once the immoralities have been located, this gives rise to other directed edges, that necessarily retain the same direction in any faithful DAG. These are known as {\em strongly protected edges} (definition~\ref{defstpred}) and the final phase of the algorithm locates these.

\section{Testing for Conditional Independence}\label{seccitest}  The divergence measure used by Tsamardinos, Brown and Aliferis in~\cite{TBA} for determining whether a conditional independence statement held is the mutual information measure, based on the Kullback Leibler divergence. That is, for two  variables $X$ and $Y$ and and a set of variables $S$, let $\hat{p}_{X|S}$, $\hat{p}_{Y|S}$ and $\hat{p}_{X,Y|S}$ denote the empirical probabilities computed from data. Then the mutual information between $X$ and $Y$ given $S = {\underline s}$ is defined as

\begin{eqnarray*} I(X,Y|S = {\underline s})&=& 2  n_S(\underline{s})\sum_{x,y} \hat{p}_{X,Y|S}(x,y|{\underline s}) \log \frac{\hat{p}_{X,Y|S}(x,y|{\underline s})}{\hat{p}_{X|S}(x|{\underline s}) \hat{p}_{Y|S}(y|{\underline s})}\\
&=& 2\sum_{x,y} n_{X,Y,S}(x,y,{\underline s})\log\frac{n_{X,Y,S}(x,y,{\underline s})n_S(\underline{s})}{n_{X,S}(x,\underline{s})n_{Y,S}(y,\underline{s})}
\end{eqnarray*}

\noindent where $\underline{s}$ denotes that the set $S$ is instantiated as $\underline{s}$, $n_{X,Y,S}(x,y,\underline s)$ denotes  the number of times that $(X,Y,S)$ is instantiated as $(x,y,\underline{s})$ in the data, similarly for $n_{X,S}(x,\underline{s})$, $n_{Y,S}(y,\underline{s})$ and $n_S(\underline{s})$. Under the hypothesis that $X \perp Y| S= {\underline s}$, 

\[ I(X,Y|S = {\underline s}) \stackrel{approx}{\sim} \chi^2_{(|X|-1)(|Y|-1)}\]

\noindent where $|X|$ and $|Y|$ denote the number of states of $X$ and $Y$ respectively. The following test statistic will be used to determine conditional independence.

\begin{Defn}[Test Statistic]\label{defteststat} The following test statistic

\begin{equation}\label{eqtestgee} G(X,Y;S) = \sum_{\underline{s}|n_{X,Y,S}(x,y,\underline{s}) \geq 5\; \forall (x,y)} I(X,Y|S = \underline{s})\end{equation}

\noindent will be used. Since the sum of independent $\chi^2$ distributions is again $\chi^2$  the corresponding number of degrees of freedom, obtained by summing,  is 

\begin{equation}\label{eqtestdf} d = N_{\geq 5} (X,Y,S)(|X|-1)(|Y|-1)
\end{equation}
where $N_{\geq 5}(X,Y,S)$ denotes the number of states of $S$ such that  $n_{X,Y,S}(x,y,\underline{s}) \geq 5$ for all $(x,y)$. If $N(X,Y,S) = 0$, then $H_1 : X \not \perp Y | S$ is not rejected.\vspace{5mm}

\noindent The critical value is given by 
\begin{equation}\label{eqtestdelta}
 \delta(X,Y,S) = \chi_d^2(1 - \alpha)  
\end{equation}

\noindent where $\alpha$ denotes the nominal significance level of $5\%$.
\end{Defn}

\noindent In~\cite{TBA}, the hypothesis of independence is {\em accepted} if $G(X,Y;S) \leq \chi_d^2(1 - \alpha)$, where $\alpha$ denotes the nominal significance level of $5\%$. There is no adjustment to accommodate conditional independence relations obtained in this way that are incompatible with independence relations that are rejected.

As indicated earlier, there are two problems with this. Firstly (and less seriously) the true significance level is clearly much higher than the nominal significance level; for large networks, the notion of `significance level' is meaningless - with any reasonable nominal significance level, and a large number of variables, it is not possible to infer that an  an independent replication of the experiment will fit the network produced.

This is not so serious, because the aim is not so much to establish a network that will be reproduced by a independent data set (which is not to be expected with large numbers of variables), but rather to find a network that fits the dependence structure exhibited in the {\em current} data set. For this, the fact that the nominal significance level is $5\%$ for each test is not a problem; it is simply a declaration of the dependence level that is necessary to be declared an `association'.

The second, and much more serious problem with this is that the procedure, is the one indicated earlier that leads to the   modification of the MMPC algorithm, for determining the conditional independence statements that are to be represented by $d$-separation statements in the graphical model. The Women and Mathematics example, on six variables, gives a situation where conditional independence statements are accepted, leading to removal of edges and $d$-separation statements in the graphical model that contradict dependence statements in the distribution. If a single hypothesis test does not reject $H_0: X \perp Y |S$, then the algorithm described by~\cite{TBA} {\em accepts} this CI statement, hence the skeleton that does not contain the edge $\langle X, Y \rangle$, even though this leads to $d$-separation statements in the graphical model where the probability distribution has dependence.

The following theorem is a consistency results, indicating that if a set of CI statements is consistent, then if $X \perp Y | W \cup Z$ and $W \perp Y|Z$ then the statement $X \perp Y|Z$ must also be in the set of conditional independence relations.

\begin{Th}\label{thci1} Let $X$ and $Y$ denote two variables and let $W,Z$ denote two disjoint sets of variables, neither containing $X$ or $Y$. If $X \perp Y | W \cup Z$ then 
\[ Y \perp W|Z \Rightarrow X \perp Y | Z.\]
 \end{Th}

\paragraph{Proof} If $X \perp Y | W \cup Z$ then, for $x,y,\underline{w}, \underline{z}$ such that $p_{X,Y,W,Z} > 0$, $p_{X,Y,Z,W} = \frac{p_{X,W,Z}p_{Y,W,Z}}{p_{W,Z}}$. It follows that  
\begin{eqnarray*} Y \perp W | Z &\Rightarrow& p_{X,Y,W,Z} = \frac{p_{X,W,Z}p_{Y,W,Z}}{p_{W,Z}}\\&& = \frac{p_{X,W,Z}p_{Y,Z}p_{W,Z}}{p_{W,Z}p_Z} = \frac{p_{X,W,Z}p_{Y,Z}}{p_Z}\\
 &\Rightarrow& X \perp Y | Z.
\end{eqnarray*}
\qed

\paragraph{The Women and Mathematics data set} This data set, discussed in more detail in subsection~\ref{subsWAM}), presents a situation with four of the variables, $B,C,D,E$ where, using a nominal significance level of $5\%$, tests give: $C \not \perp E$, $C \not \perp E|D$, but $C \perp B$ and $C \perp D$ are not rejected. The chi squared test applied to $H_0: C \perp E | \{B,D\}$ also fails to reject the null hypothesis.

But, by theorem~\ref{thci1}, if $E \perp C|\{B,D\}$, then $\{B,D\} \perp C \Rightarrow C \perp E$, while the result of the hypothesis test is: reject $C \perp E$ and accept $C \not \perp E$. This is a specific example where accepting conditional independence when the result of the hypothesis test is `insufficient information to reject conditional independence' leads to a graphical model that does not represent all the associations between the variables. Since the hypothesis $E \perp C | \{B,D\}$ is not rejected, the final skeleton from the MMPC algorithm does not contain an edge between $E$ and $C$. Neither does it contain edges $\langle E,B \rangle$ or $\langle E,D \rangle$.  Therefore, in the corresponding graphical model for variables $B,C,D,E$, $E$ is $d$-separated from $C$, even though $E \not \perp C$. \qed \vspace{5mm}

\noindent The procedure to correct this uses three results; theorem~\ref{thci1} above and theorems~\ref{thci2} and~\ref{thci3} below.

\begin{Th}\label{thci2}
Let $V = \{X_1, \ldots, X_d\}$ denote a set of $d$ variables.
Let $S = \{X_{k_1}, \ldots, X_{k_m}\} \subset V$. Then, for $X_i, X_j \not \in S$, if $X_i \perp X_j | S$ 
\[X_i \perp X_{k_m} |  \{X_{k_1}, \ldots, X_{k_{m-1}}, X_j\}  \Leftrightarrow X_i \perp \{X_{k_m}, X_j\}.\]
\end{Th}

\paragraph{Proof of theorem~\ref{thci2}} Let $R = \{X_{k_1}, \ldots, X_{k_{m-1}}\}$ and assume that $X_i \perp X_j | S$. Recall that
\[X_i \perp X_j | S \Leftrightarrow p_{X_i, X_j, S}(x_i, x_j, {\underline s}) = \left\{\begin{array}{ll} \frac{p_{X_i,S}(x_i, {\underline s})p_{X_j,S}(x_j, {\underline s})}{p_S({\underline s})} & p_S({\underline s}) > 0 \\ 0 & p_S({\underline s}) =0 \end{array}\right. \]

\noindent If  \[ p_{X_i, X_j S} = \frac{p_{X_i,S}p_{X_j,S}}{p_S}\] \noindent for all ${\underline s}$ such that $p_{S}({\underline s}) > 0$, then, for all $\underline{s}$ such that $p_{S}( \underline{s}) > 0$, it follows that   
\[ p_{X_i,X_j,S} = \frac{p_{X_i, X_j, R}p_{X_j, S}}{p_{R\cup\{X_j\}}} 
  \Leftrightarrow   p_{X_i|S} = p_{X_i|R \cup \{X_j\}} 
\]
\noindent so that 
\[ X_i \perp X_{k_m} |  \{X_{k_1}, \ldots, X_{k_{m-1}}, X_j\} \Leftrightarrow X_i \perp \{X_j, X_{k_m}\} .\]
\noindent as required. 
\qed 

\begin{Th}\label{thci3}  Let $(B(i,j))_{1 \leq i < j \leq k}$ be a collection of numbers such that $B(i,j) = 1$ or $0$ for each $(i,j)$. For any such collection $(B(i,j))_{1 \leq i < j \leq k}$, there is a collection of random variables $\{X_1, \ldots, X_k\}$ such that $X_i \perp X_j$ if $B(i,j) = 0$ and $X_i \not \perp X_j$ if $B(i,j) = 1$. 
\end{Th}

\paragraph{Proof of theorem~\ref{thci3}} The proof consists of showing that there exists a Bayesian network, that may contain more than $k$ variables, such that $X_i$ and $X_j$ are $d$-separated by the empty set if $B(i,j) = 0$ and where $X_i$ and $X_j$ are $d$-connected (when all other variables are not instantiated) when $B(i,j) = 1$. 

Consider a directed acyclic graph constructed in the following way. Start with a graph containing nodes $\{X_1, \ldots, X_k\}$ and no edges in the graph. Whenever $B(i,j) = 1$, add a node $T_{i,j}$ and directed arrows $T_{ij} \mapsto X_i$ and $T_{ij} \mapsto X_j$. For $B(i,j) = 0$ do not add any additional nodes and do not add any directed arrows to the graph. 

For the graph constructed in this way, ${\cal G} = (W, D)$ where $W$ is the node set and $D$ is the directed edge set, consider any distribution $p$ for which ${\cal G} = (W,D)$ is a faithful graphical model. It is possible to construct such a probability distribution over the variables in $W$. Then, since the original variables $\{X_1, \ldots, X_k\}$ only appear in collider connections, and since an additional variable  $T_{ij}$ only appears in fork connection between $X_i$ and $X_j$, it follows that $X_i \not \perp X_j \Leftrightarrow B(i,j) = 1$.   Now  marginalise over the variables $(T_{ij})_{(i,j): B(i,j) = 1}$ and the distribution thus obtained $p_{X_1,\ldots, X_k}$ satisfies the desired property. 
\qed

\subsection{Determining Conditional Independence: The Procedure}\label{subsciproc} The $M^3$PC algorithm (modified MMPC algorithm) modifies the procedure to decide whether a conditional independence statement is to be included, in the following way.

With an initialisation of $A(X,Y,S) = 2$ for each $(X,Y,S)$ before the independence relations have been established, let $A(X,Y;S) = 1$ denote that $X \not \perp Y|S$ is to be included in the set of dependence relations and let $A(X,Y;S) = 0$ denote that $X \perp Y | S$ is to be included in the set of dependence relations. Once a value of $0$ or $1$ has been established, it does not change for the remainder of the procedure. Those which do not have a value of $0$ or $1$ by the end of the procedure are not used to determine which edges the graph contains. 

 The procedure computes values of $A(X,Y;S)$ starting with $S= \phi$ and only computing $A(X,Y;S)$ once $A(X,Y;R)$ has been established for all $R \subset S$ (strict subsets). 

\label{page} For two variables $X$ and $Y$ and a subset $S$, the value $A(X,Y;S) = 0$ is established if and only if the following three criteria are satisfied:

\begin{enumerate} \item $G(X,Y|S) \leq \delta(X,Y;S)$, where $G(X,Y|S)$ and $\delta(X,Y;S)$ are defined by equation~\ref{eqtestgee}. 
\item for each  $Z \in S$, 

\[ A(Y,Z; S\backslash \{Z\}) = 0 \Rightarrow  A(X,Y; S \backslash \{Z\}) = 0  \]

\noindent and

\[A(X,Z; S \backslash \{Z\}) = 0 \Rightarrow  A(X,Y; S \backslash \{Z\}) = 0. \]

\item For each $Z \in S$,
\[A(X,Z;\{Y\}\cup S \backslash\{Z\}) = 0 \Leftrightarrow X \perp \{Y,Z\}\]
\noindent and
\[A(Y,Z;\{X\}\cup S \backslash \{Z\}) = 0 \Leftrightarrow Y \perp \{X,Z\}.\]

\end{enumerate}  If either of the first two criteria is not satisfied, the value $A(X,Y;S) = 1$ is returned. If the first two are satisfied, but the third is not satisfied, then if $X \not \perp \{Y,Z\}$, but $A(X,Z; \{Y\} \cup S \backslash \{Z\}) = 0$, then set  $A(X,Y;S)= 1$. \vspace{5mm}

\noindent The value $A(X,Y;S) = 0$ is returned if and only if all three criteria are satisfied.\vspace{5mm}

\noindent This could be modified in the following way: if $X \not \perp \{Y,Z\}$, but $A(X,Z; \{Y\} \cup S \backslash \{Z\}) = 0$, then set $A(X,Z; \{Y\} \cup S \backslash \{Z\}) = 1$ if $G(X,Z;\{Y\} \cup S \backslash \{Z\}) > G(X,Y ; S)$. If  $G(X,Z;\{Y\} \cup S \backslash \{Z\}) \leq  G(X,Y ; S)$, then set  $A(X,Y;S)= 1$.  

If $Y \not \perp \{X,Z\}$, but $A(Y,Z; \{X\} \cup S \backslash \{Z\}) = 0$, then set $A(Y,Z; \{X\} \cup S \backslash \{Z\}) = 1$ if $G(Y,Z;\{X\} \cup S \backslash \{Z\}) > G(X,Y ; S)$. If  $G(Y,Z;\{X\} \cup S \backslash \{Z\}) \leq  G(X,Y ; S)$, set $A(X,Y;S)= 1$. 

\begin{Th} The statements $A(X,Y;S)$ decided according to the procedure given above, will be consistent with each other. That is, there exists a collection of random variables such that for any pair of variables $X,Y$ and a subset $S$, $X \not \perp Y|S$ when $A(X,Y;S) = 1$ and $X \perp Y|S$ when $A(X,Y;S) = 0$.  
\end{Th}

\noindent {\bf Proof} The proof is by induction on the number of variables permitted in the conditioning set. By theorem~\ref{thci3}, no inconsistencies are introduced by the statements where the conditioning set is $\phi$.

Assume that there are no inconsistencies when all the values $A(X,Y;S)$ are considered for all $(X,Y,S)$ such that $X,Y \not \in S$, and  $|S| \leq n$, where $|S|$ denotes the number of variables in $S$. A statement $A(X,Y;S)$ where $|S| = n$ will be referred to as a statement of order $n$. 

Now consider a statement $A(X,Y;S)$, where $|S| = n+1$, derived according to the principles listed above. If $A(X,Y;S) = 1$, then it is not inconsistent with any previous statements. If $A(X,Y,S) = 0$, then  

\begin{equation}\label{eqndec} A(X,Y;S) = 0 \quad \Leftrightarrow \quad p_{X,Y,S} = \frac{p_{X,S}p_{Y,S}}{p_S} \quad \forall {\underline s} : p_S({\underline s}) > 0\end{equation}

\noindent It always holds that $p_{X,Y,S} = p_{X|Y,S}p_{Y,S} = p_{Y|X,S}p_{X,S}$. If $|S| = n+1$, then there is a conditional independence relation of order $n+1$ between the variables in the set $\{X,Y\} \cup S$ if and only if either 

\[p_{X,Y,S} = p_{X|R}p_{Y,S} \qquad \mbox{for some}\; R \subset S \cup \{Y\}\;\mbox{or}   \qquad p_{X,Y,S} = p_{X,S}p_{Y|R}\qquad  \mbox{for some}\;  R \subset S \cup \{X\}. \]

\noindent Without loss of generality, only consider the first of these; $p_{X,Y,S} = p_{X|R}p_{Y,S}$ where $R$ is a strict subset of $S \cup \{Y\}$, since the second is similar.   Let $S = \{Z_1,\ldots, Z_{n+1}\}$. Then $p_{X,Y,S} = p_{X|R}p_{Y,S}$ where $R \subseteq \{Y\}\cup S \backslash \{Z_i\}$ for some $i \in \{1, \ldots, n+1\}$ if and only if $A(X,Z_i; \{Y\}\cup S \backslash \{Z_i\}) = 0$. Either $Y \in R$ or $Y \not \in R$. Suppose $Y \in R$. Then   if $X \perp Y | S$, it follows that 
\[X \perp Z_i | \{Y\} \cup S \backslash \{Z_i\} \Leftrightarrow p_{X|Y,S} = p_{X|Y,S\backslash\{Z_i\}} = p_{X|S} = p_{X|S\backslash\{Z_i\}}\]
\noindent so that
\[X \perp Z_i | \{Y\} \cup S \backslash \{Z_i\} \Leftrightarrow X \perp \{Y,Z_i\}| S \backslash \{Z_i\}.\]

\noindent Suppose $Y \not \in R$. Then if $X \perp Y|S$ it follows that 
\[ X \perp Z_i | S \backslash \{Z_i\} \Leftrightarrow p_{X,Y,S} = \frac{p_{X,S\backslash \{Z_i\}}p_{S}p_{Y,S}}{p_{S \backslash \{Z_i\}}p_S} = \frac{p_{X,S \backslash \{Z_i\}}p_{S,Y}}{p_{S \backslash \{Z_i\}}} \Leftrightarrow X \perp \{Y,Z_i\} | S \backslash \{Z_i\}.\]

\noindent From the construction method, if the value $A(X, Z_i ; \{Y\} \cup S \backslash \{Z_i\}) = 0$ is already assigned, then a necessary condition for $A(X,Y;S) = 0$ is that the values $A(X,Y; S \backslash \{Z_i\})= 0$ and $A(X, Z_i; S \backslash\{Z_i\}) = 0$ have already been assigned. 

If the value $A(X,Z_i; S \backslash \{Z_i\}) = 0$ has been assigned, then a necessary condition for $A(X,Y; S) = 0$ is that the value $A(X,Y; S \backslash \{Z_i\}) = 0$ has already been assigned.

The construction is such that all the conditional independence statements accepted are consistent with the conditional independence / dependence statements of a lower or equal order.\qed \vspace{5mm}

\noindent Therefore, if the procedure outlined above is adopted,  the set of conditional independence statements that are accepted will be logically consistent with the set of conditional independence statements that are rejected in the hypothesis tests.

\section{The Modified Maximum Minimum Parents and Children Algorithm} 

This section gives the Maximum Minimum Parents and Children (MMPC) algorithm presented in~\cite{TBA}, followed by the modified Maximum Minimum Parents and Children ($M^3$PC) algorithm that is the main contribution of this article. The original MMPC algorithm is presented in a way that makes it clear, when compared with the $M^3$PC algorithm,  where the modification the locates the immoralities occurs.  

The $M^3$PC algorithm runs in two stages: 
\begin{enumerate} \item run the MMPC algorithm, with the modifications required to obtain both the skeleton and the immoralities,
\item locate the additional edges that are strongly protected. 
\end{enumerate}

Most of the work is in the first part; locating the protected edges after the immoralities have been located is relatively straightforward. The criterion for assessing the computational efficiency of an algorithm is the number of {\em statistical calls}  that have to be carried out. A statistical call is defined as follows.

\begin{Defn}[Statistical Call]\label{defstatcall} A {\em statistical call} is a   computation, for example, computing the value of $G(X_i, X_j; S)$ (definition~\ref{defteststat}, equation(\ref{eqtestgee})), that measures a level of statistical association.
 \end{Defn} 

\noindent This has become the standard measure of efficiency, following the criteria stated in~\cite{TBA} section 7, which followed Spirtes, Glymour and Scheines~\cite{SGS} and Margritis and Thrun~\cite{MT}. The location of strongly protected edges after the immoralities have been obtained does not increase the order of the computational time. When the data set is large (say order 5000 instantiations of 6000 variables), and the statistical call involves computing the marginal empirical probabilities, then the time taken for a statistical call is essentially the time taken to compute the cell counts. For example, if there are 5000 observations and all the variables are binary, then since the tests require counts of greater than or equal to $5$, the  largest marginal that can be considered has less than $1000$ cells, corresponding to less than $10$  variables, since $2^{10} = 1024$. 

\subsection{The  Maximum Minimum Parents and Children  Algorithm}
The variable set is denoted by $V = \{X_1, \ldots, X_d\}$. The MMPC algorithm introduced in~\cite{TBA} is the combination of   algorithms~\ref{MMPCalg1},~\ref{MMPCalg2} and~\ref{MMPCalg3}; run in the following sequence: stage 1: algorithm~\ref{MMPCalg1}, stage 2: algorithm~\ref{MMPCalg2}, stage 3: algorithm~\ref{MMPCalg3}, stage 4: algorithm~\ref{MMPCalg2}.

\paragraph{Stage 1} Each variable in turn, for $j = 1,\ldots, d$, is taken as the `target variable' (to use the notation of~\cite{TBA}). The first stage (algorithm~\ref{MMPCalg1}) produces for each variable $X_j$ a set ${\cal Z}_j$ which contains all the neighbours of $X_j$ in the final graph, along with other variables that will be eliminated in later stages.

\paragraph{Stage 2} The second stage (algorithm~\ref{MMPCalg2}) simply observes that if a variable $X_k$ is not in the neighbour set of $X_j$, then $X_j$ cannot be in the neighbour set of $X_k$, so that if $X_k \not \in {\cal Z}_j$, then $X_j$ should be removed from ${\cal Z}_k$. This stage of the algorithm is not computationally expensive, since it is purely algebraic and there are no statistical calls. This run of algorithm~\ref{MMPCalg2} is unnecessary, but improves computational efficiency; it reduces the numbers and sizes of subsets that have to be considered in the next stage.

\paragraph{Stage 3} The third stage of the algorithm (algorithm~\ref{MMPCalg3}) may be computationally   expensive if the sizes of the sets produced from stage 1 and stage 2 are large, since it can require searching through rather large collections of subsets. 

Starting with ${\cal Z}_j$ as the set of possible neighbours of $X_j$ after stage 2, let ${\cal Y}_{j,k}$ denote the set of possible neighbours of $X_j$ after variables $X_1,\ldots, X_k$ have been considered, where $X_k$ has been eliminated if it was present in ${\cal Y}_{j,k-1}$ and  if a subset $S$ of ${\cal Y}_{j,k-1}$ was located such that $X_j \perp X_k | S$. Consider a particular DAG that is faithful to the distribution. Clearly, $X_k$ cannot be both an ancestor and a descendant of $X_j$ in the DAG, otherwise the DAG contains a cycle. If $X_k$ is an ancestor of $X_j$, then since $\Pi_j \subseteq S$, taking $S = \Pi_j$ will give a subset such that $X_j \perp X_k | S$, since all trails from $X_k$ to $X_j$ will have an instantiated chain or fork connection (if the trail passes through a parent of $X_j$) or an uninstantiated collider connection. Clearly, at least one of these necessarily occurs if $X_k$ is an ancestor of $X_j$ in a DAG.

If $X_k$ is a descendant of $X_j$ in the DAG, then it is possible that $X_k$ may not be eliminated from ${\cal Y}_{j,k-1}$, even if there is no directed edge $(X_j, X_k)$ (that is, from $X_j$ to $X_k$) in the DAG. This happens if and only if on every trail from $X_j$ to $X_k$, every chain node on the trail that is a neighbour of $X_j$, required to block the trail when only variables from the neighbour set of $X_j$ are admitted, is also a collider node on a different trail from $X_j$ to $X_k$, and the variable $Z$ such that there is a collider connection $X_j \rightarrow Y \leftarrow Z$ between $X_j$ and $Z$ is not in ${\cal Z}_j$.  

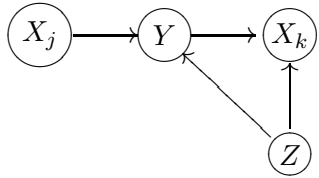
\begin{figure} 
\begin{center} 
\[ \UseTips \xymatrix{ 
*++[o][F]{X_j}\ar[r] & *++[o][F]{Y}\ar[r] &   *++[o][F]{X_k}  \\
  &   &   *++[o][F]{Z}\ar[ul] \ar[u]  } \]
\end{center}
\caption{$X_k$ not eliminated from ${\cal Z}_j$}\label{MMPCst2} 
\end{figure}

The   configuration in figure~\ref{MMPCst2} represents such a situation; there are two trails from $X_j$ to $X_k$, $Y$ is a chain node in one and a collider node in the other. It is required that $Y$ is instantiated if the trail $X_j -Y-X_k$ is to be blocked by instantiating nodes from ${\cal Z}_j$. Suppose that, after stage 2, $Z \not \in {\cal Z}_j$, but $\{Y, X_k\} \in {\cal Z}_j$. If the variable $Y$ is instantiated, then there is an open trail $X_k - Z - Y - X_j$. If $Y$ is not instantiated, then there is an open trail $X_k - Y - X_j$. In both cases, $X_j$ and $X_k$ are $d$-connected.

In the example in figure~\ref{MMPCst2}, although $X_k$ is not eliminated from ${\cal Z}_j$ at this stage, the variable $X_j$ will be eliminated from ${\cal Z}_k$, since $X_j \perp X_k \|_{{\cal G}} \{Y,Z\}$. 

In general, following the discussion above, by instantiating $\Pi_k$, the parents of $X_k$, $X_j$ will be removed from ${\cal Z}_k$. 

For a particular faithful DAG ${\cal G}$, if $X_k$ is neither a parent nor a child of $X_j$ in ${\cal G}$, then take $S = \Pi_j$. Then every trail between $X_j$ and $X_k$ contains either an instantiated fork or chain (if the trail passes through a parent of $X_j$) or an uninstantiated collider (if it does not pass through a parent of $X_j$) and hence $X_k$ is eliminated from $X_j$ at this stage. 

\paragraph{Stage 4} Since, after stage 3, there is the possibility of neighbours that are not present in the skeleton of a faithful DAG, it is  necessary and sufficient to run algorithm~\ref{MMPCalg2} again. After stages 1 - 4, the resulting graph obtained is the skeleton of a faithful graph; it is clear that if ${\cal Z}_1, \ldots, {\cal Z}_d$ are the neighbour sets for $X_1, \ldots, X_d$ respectively, then there is an  edge $\langle X,Y \rangle$ in the skeleton   if and only if  there does not exist a set $S$ such that $X \perp Y |S$. Following theorem~\ref{thfaithedge}, it is therefore the skeleton of a faithful graph. 
 
\begin{algorithm}
\caption{The MMPC Algorithm, Stage 1}
\label{MMPCalg1}
\begin{algorithmic}
\FOR{$j = 1, \ldots, d$ } 
 \STATE Set ${\cal Z}_{j,0} = \phi$ (the empty set)
 \FOR{$i = 1,\ldots, d$}
 \IF{$i = j$} \STATE Set ${\cal Z}_{j,i} = {\cal Z}_{j,i-1}$
 \ELSE
 \IF {$X_i \perp X_j | {\cal Z}_{j,i-1}$} \STATE Set ${\cal Z}_{j,i} = {\cal Z}_{j,i-1}$
\ELSE \STATE Set ${\cal Z}_{j,i} = {\cal Z}_{j, i-1} \cup \{X_i\}$
  \ENDIF
  \ENDIF
 \ENDFOR
\STATE Set ${\cal Z}_j = {\cal Z}_{j,d}$
\ENDFOR  
\end{algorithmic}
\end{algorithm}

\begin{algorithm}
\caption{The MMPC Algorithm, Stages 2 and 4}
\label{MMPCalg2}
\begin{algorithmic}
\FOR{$j = 1,\ldots, d$} \STATE Set ${\cal Y}_{j,0} = {\cal Z}_j$.
\FOR{$k = 1,\ldots, d$}
\IF{$X_k \in {\cal Y}_{j,k-1}$ but $X_j \not \in {\cal Z}_k$}
\STATE Set ${\cal Y}_{j,k} = {\cal Y}_{j,k-1} \backslash \{X_k\}$
\ELSE \STATE Set ${\cal Y}_{j,k} = {\cal Y}_{j,k-1}$
\ENDIF
\ENDFOR
\STATE Set ${\cal Z}_j = {\cal Y}_{j,d}$
\ENDFOR
\end{algorithmic}
\end{algorithm}

\begin{algorithm}
\caption{The MMPC Algorithm, Stage 3}
\label{MMPCalg3}
\begin{algorithmic}
\FOR{$j = 1, \ldots, d$} \STATE Set ${\cal Y}_{j,0} = {\cal Z}_j$.
\FOR{$k = 1, \ldots, d$,}
\IF{$X_k \in {\cal Y}_{j,k-1}$ and $\exists S \subseteq {\cal Y}_{j,k-1} \backslash \{X_k\}$ such that $X_j \perp X_k | S$} \STATE Set ${\cal Y}_{j,k} = {\cal Z}_{j,k-1} \backslash \{X_k\}$.
\ELSE \STATE Set ${\cal Y}_{j,k} = {\cal Y}_{j,k-1}$.
\ENDIF
\ENDFOR
\STATE Set ${\cal Z}_j = {\cal Y}_{j,d}$
\ENDFOR
 \end{algorithmic}
\end{algorithm}

\subsection{The Modified MMPC Algorithm}
 
The modification of the MMPC algorithm to obtain the skeleton and the immoralities is presented in algorithms ~\ref{MMPCmodalg1}, ~\ref{MMPCmodalg2}, ~\ref{MMPCmodalg3}, ~\ref{MMPCmodalg4}, ~\ref{MMPCmodalg5} and ~\ref{MMPCmodalg6}, with algorithm~\ref{MMPCmodalg0} used to obtain conditional independence / dependence relations that are consistent with each other.  The additional strongly protected edges (definition~\ref{defstpred}), whose directions are forced after the immoralities are determined, are located by algorithms~\ref{MMPCstpred1},~\ref{MMPCstpred2} and~\ref{MMPCstpred3}, carried out in that sequence. The $M^3PC$ algorithm is therefore the sequence: algorithms~\ref{MMPCmodalg1},~\ref{MMPCmodalg2},~\ref{MMPCmodalg3},~\ref{MMPCmodalg4},~\ref{MMPCmodalg5},~\ref{MMPCmodalg6},~\ref{MMPCstpred1},~\ref{MMPCstpred2},~\ref{MMPCstpred3}, run in that order, followed by algorithm~\ref{MMPCcheck} to ensure that a valid graphical model is returned even if the dependence / independence relations returned do not correspond to a probability distribution that has a faithful graphical model.

Algorithm~\ref{MMPCmodalg0} is used whenever a conditional independence / dependence relation has to be established. 

The way that the modification for locating the immoralities to the MMPC algorithm is made,  can be seen from the listings of algorithms~\ref{MMPCmodalg1} - ~\ref{MMPCmodalg5} and comparing them with the corresponding routines for the MMPC algorithm.   They are  described below, comparing the various stages with the MMPC algorithm,  showing that stages 1 - 5 of the algorithm return the correct skeleton and the immoralities and pinpointing  where the original MMPC algorithm has been modified. 

Stage 6, which is listed as algorithm~\ref{MMPCmodalg6},  constructs the sets of edges that are, at this stage directed and the remaining edges which are undirected. Stages 7,8 and 9 (algorithms~\ref{MMPCstpred1}, ~\ref{MMPCstpred2} and ~\ref{MMPCstpred3}) then locate the remaining strongly protected edges (definition~\ref{defstpred}), so that the resulting graph is the essential graph.

At this stage,  the algorithm would usually be finished, since (as discussed in~\cite{}), underlying probability distributions usually have a faithful graphical model. The problem is that, when establishing dependence from data, a dependence structure in the original distribution will not be accepted into the model if there is insufficient evidence from the data. This can lead to a set of dependence / independence relations that do not correspond to a probability distribution with a faithful graphical representation. This can lead to situations where the algorithm, after stage 9, returns a graph that is not the essential graph corresponding to the directed acyclic graph of any Bayesian network. Stage 10 (algorithm~\ref{MMPCcheck}) deals with this and returns an essential graph.

If it is not necessary to employ algorithm~\ref{MMPCcheck}, then the graphical model returned is similar in quality to the network returned by the MMHC algorithm of~\cite{TBA}. If algorithm~\ref{MMPCcheck} is necessary, the network returned may fit the data substantially less well than the network returned by the MMHC algorithm.  

The definition of parents, children, ancestors and descendants is given in definition~\ref{defpcad}. Then a variable $X$ is $d$-separated (definition~\ref{defdsep}) from the rest of the network by the {\em Markov blanket} $M(X)$ (definition~\ref{defMB}), which is the set of parents and children and parents of children of $X$, from all the other variables in the network (exercise 6 on page 77 of~\cite{KN}). For a faithful graph, $d$-separation and conditional independence statements are equivalent. Theorem~\ref{thmeq} shows that $d$ separation implies conditional independence; by the definition of faithfulness (definition~\ref{deffaith}), $d$ separation and conditional independence are equivalent for a faithful graph. Theorem~\ref{thfaithedge} is crucial in the justification of the MMPC algorithm; it states that if a graph ${\cal G}$ is faithful to a probability distribution $p$ then there is an edge $\langle X_i, X_j \rangle$ in the skeleton if and only if there does not exist a subset  $S$ of variables such that $X_i \perp X_j | S$. 

Locating the immoralities is a minor extension of this principle; if ${\cal G}$ is a faithful graph and there is a vee structure (definition~\ref{defveestr}) $X-Y-Z$ in the skeleton, and $S$ is a set such that $X \perp Z | S$ with $Y \not \in S$, then $X-Y-Z$ is an immorality. This is the content of theorem~\ref{thveeimm}. 

In computation, it may be safer to require also that, if $X-Y-Z$ is a vee-structure in the skeleton, then it is an immorality if and only if there exists a set $S$ with $Y \not \in S$ such that $X \perp Z | S$ and also $X \not \perp Z | S \cup \{Y\}$. \vspace{5mm}

\noindent With these principles for constructing the skeleton and locating the immoralities,  the justification proceeds as follows.

\paragraph{Conditional Independence Tests} Algorithm~\ref{MMPCmodalg0} is used to determine the conditional independence structure used to construct the skeleton and immoralities of a graphical model. 

For a subset $S = \{X_{k_1}, \ldots, X_{k_m}\} \subseteq \{X_1, \ldots, X_d\}$, the notation $A(i,j;k_1,\ldots, k_m)$ and $A(i,j;S)$ will be used interchangeably. 

The initialisation is $A(i,j;S) = 2$ (or some value different from $0$ or $1$) for all $(i,j;S)$.

A value $A(i,j;S) = 0$ is established if and only if the relation $X_i \perp X_j|S$ is added to the conditional independence relations and a value $A(i,j;S) = 1$ is established if and only if $X_i \perp X_j|S$ is rejected. 

Once a value of $A(i,j;S)$ is determined, then it is stored and it is not altered.

Algorithm~\ref{MMPCmodalg0} determines $A(k_1,k_2;k_3, \ldots, k_{m+2})$, according to the three principles listed on page~\pageref{page}. In the listing of algorithm~\ref{MMPCmodalg0}, the following condition is referred to. 

\begin{Defn}[Condition $\Xi(k_1,k_2;k_3,\ldots,k_{i+1})$]~\label{defcondxi} For a subset of variables \[ \{X_{k_1}, \ldots, X_{k_{i+1}}\} \subseteq \{X_1, \ldots, X_d\},\] condition $\Xi(k_1,k_2;k_3,\ldots,k_{i+1})$ is said to hold if and only if the following three conditions hold:
\begin{itemize}
 \item 
\[ G(X_{k_1}, X_{k_2}|X_{k_3}, \ldots, X_{k_{i+2}}) < \delta(X_{k_1}, X_{k_2}|X_{k_3}, \ldots, X_{k_{i+2}})\]
where $G$ and $\delta$ are given in definition~\ref{defteststat}, equations (\ref{eqtestgee}) and (\ref{eqtestdelta}).
\item For $j = 3, \ldots, i+2$, 
\[ A(k_{1}, k_{j}; \{k_{3}, \ldots, k_{i+2}\}\backslash \{k_{j}\}) = 0 \Rightarrow A(k_{1}, k_{2}; \{k_{3}, \ldots, k_{i+2}\}\backslash \{k_{j}\}) = 0\]
\item For $j = 3, \ldots, i+2$, 
\[A(k_{2}, k_{j}; \{k_{3}, \ldots, k_{i+2}\}\backslash \{k_{j}\})=0 \Rightarrow A(k_{1}, k_{2} ; \{k_{3}, \ldots, k_{i+2}\}\backslash \{k_{j}\})=0\]
\end{itemize}
\end{Defn}

\paragraph{Stage 1} This stage of the $M^3$PC algorithm is algorithm~\ref{MMPCmodalg1} and corresponds to algorithm~\ref{MMPCalg1} of the MMPC algorithm. It differs from algorithm~\ref{MMPCalg1} (stage 1 of the MMPC algorithm) in that whenever it is decided to remove an edge $\langle X_i, X_j \rangle$ because $X_i \perp X_j|S_{ij}$, the set $S_{ij}$ is recorded. This is because if $X_i \perp X_j |S_{ij}$, then this may be used to determine whether or not any vee - structures (definition~\ref{defveestr}) $X_i - Y - X_j$ are immoralities in stage 2, listed as algorithm~\ref{MMPCmodalg2}. 

A conditional independence statement $X_j \perp X_k | S$ implies that $X_k$ is not a neighbour of $X_j$. The neighbours of $X_1, \ldots, X_d$ are therefore subsets of ${\cal Z}_1, \ldots, {\cal Z}_d$ respectively. 

\paragraph{Stage 2} This stage in the $M^3$PC algorithm is algorithm~\ref{MMPCmodalg2}. This is a modification of algorithm~\ref{MMPCalg2}, where the following major component is added: if $X_j \not \in {\cal Z}_k$, then for each variable $Y \in {\cal Z}_j \cap {\cal Z}_k$, it is checked whether or the triple  $(X_j, Y, X_k)$ will be an immorality if $\langle X_j, Y \rangle$ and $\langle Y, X_k \rangle$ remain in the skeleton at the end. This is straightforward following the additional storage from stage 1; by definition, if $X_j$ is excluded from ${\cal Z}_k$, it follows that $X_j \perp X_k|S_{jk}$. It therefore follows from theorem~\ref{thveeimm} that if both $\langle X_j, Y \rangle$ and $\langle Y, X_k \rangle$ are in the skeleton, then $(X_j,Y,X_k)$ is an immorality if and only if $Y \not \in S_{jk}$. 

This is a necessary stage in the $M^3PC$ algorithm, unlike stage 2 (listed as algorithm~\ref{MMPCalg2}) of the MMPC algorithm. 

If $X_j \in {\cal Z}_k$ but $X_k \not \in {\cal Z}_j$, then the statement ${\cal Y}_{j,k} = {\cal Y}_{j,k-1} \backslash \{X_k\}$ removes variable $X_k$ from the possible neighbour set of $X_j$, just as in stage 2 of the MMPC algorithm.

At the end of stage 2, all structures between three variables that are vee - structures $X-Y-Z$ using the current neighbour sets,  have been checked; it has been determined whether or not they are immoralities, should the corresponding vee - structures still be present in the final graph.

\paragraph{Stage 3} This stage of the $M^3PC$ algorithm is listed as algorithm~\ref{MMPCmodalg3}. This is a modification of algorithm~\ref{MMPCalg3}, with the following important addition. When a variable is removed, this may create additional vee structures and it has to be determined whether or not these will be  immoralities, should they still be present in the final graph. 

The procedure for deciding when a variable is to be removed from ${\cal Z}_j$ is the same as that in algorithm~\ref{MMPCalg3}, but when it is decided to remove an edge $X_i$ from $X_j$, the triples  $(X_i, Y , X_j )$ are investigated for all $Y \in {\cal Z}_i \cap {\cal Z}_j$ (where ${\cal Z}_i$ and ${\cal Z}_j$ are the current candidate neighbour sets for $X_i$ and $X_j$ respectively) and tested to see whether or not $(X_i, Y, X_j)$ is an immorality. The computationally expensive part of this is the part that is used in the MMPC algorithm of~\cite{TBA}. The modification does not introduce substantial additional costs. 

Once one subset $S$ has been established such that $X_i \perp X_j | S$ (the same set as the MMPC algorithm uses), it only has to be determined whether or not $Y \in S$. If $Y \not \in S$, then $(X_i, Y, X_j)$ is  an immorality provided $X_i-Y-X_j$ is a vee structure in the final skeleton. If $Y \in S$ then $(X_i, Y, X_j)$ is not an immorality.

\paragraph{Stage 4} This stage of the $M^3PC$ algorithm is  algorithm~\ref{MMPCmodalg4} and is exactly the same as stage 4 of the MMPC algorithm, algorithm~\ref{MMPCalg2}. It is inserted for exactly the same reason described in stage 4 of the MMPC algorithm; there are situations (described in the discussion of stage 4 of the MMPC algorithm) where at this stage $X_j \in {\cal Z}_k$, but $X_k \not \in {\cal Z}_j$. The node $X_j$ has to be removed from ${\cal Z}_k$. Since this does not alter ${\cal Z}_j \cap {\cal Z}_k$, no new vee structures are created.

\paragraph{Stage 5} This stage of the $M^3PC$ considers the set of triples ${\cal I}$ that are candidates for immoralities. Any triple  $(X,Y,Z) \in {\cal I}$ is an immorality if and only if   both $\langle X, Y \rangle$ and $\langle Y, Z \rangle$ are in the skeleton. This stage is listed as algorithm~\ref{MMPCmodalg5}. The first part of this stage removes those where edges have been deleted, while the second part of algorithm~\ref{MMPCmodalg5} orders the remaining members of ${\cal I}$ so that they are listed as $(X_i, Y X_j)$ for $i < j$.\vspace{5mm}

\noindent If the independence / dependence relations correspond to those of  a probability distribution that has a faithful graphical representation, then there will not arise a situation where both $(X,Y,Z)$ and $(Y,Z,W)$ are in the set of immoralities. This would lead to the direction $Z \mapsto Y$ for the direction of the directed edge between $Y$ and $Z$ from the first of these and $Y \mapsto Z$ for the direction of the directed edge between $Y$ and $Z$ for the second of these. 

The problem is that, even if the underlying distribution has a faithful graphical representation, a conditional independence relation may be inferred from the data. If the sample size is large enough, then this will happen with small probability for any given relation, but since large graphs are in view, this situation may happen and has to be considered. The final stage of algorithm~\ref{MMPCmodalg5} deals with this, carrying out additional tests if necessary when there are contradicting immoralities. The triple $(X,Y,Z)$ is in ${\cal I}$ provided it is a vee - structure and also there exists a set $S$, such that $Y \not \in S$ and $X \perp Z | S$. If it is an immorality, it should also satisfy $X \perp Z | S \cup \{Y\}$.  

\paragraph{Summary of stages 1 - 5} After stages 1 - 5 have been complete, the skeleton is returned for a graphical model that correctly represents the dependence / independence relations established, provided there is a faithful graph that fits them.  Given the dependence / independence relations, the parts of the algorithm devoted to producing the skeleton are the same as the MMPC algorithm. 

In addition, for every vee structure $X-Y-Z$ in the graph, it has been determined whether or not $(X,Y,Z)$ is an immorality. No additional statistical calls are required; it is simply required to test whether the variable $Y$ is in the set $S$ that the $M^3PC$ algorithm uses to exclude the edge between $X$ and $Z$; the edge $\langle X,Z \rangle$ is excluded because $X \perp Z | S$ and $X-Y-Z$ is an immorality if $Y \not \in S$.  

\paragraph{Stage 6} Stage 6 is listed as algorithm~\ref{MMPCmodalg6}. It simply constructs the sets of directed and undirected edges from the set ${\cal I}$ of immoralities and the sets ${\cal Z}_1, \ldots, {\cal Z}_d$, of neighbours of $X_1, \ldots, X_d$ respectively.

\paragraph{Stages 7, 8, 9} The edge set $E$ of an essential graph may be decomposed as $E = D \cup U$ where $D$ is the set of directed edges and $U$ the set of undirected edges. Lemma~\ref{stprjust} states that the structures in figure~\ref{MMPCst2} give all the situations where the direction of an edge $(W,Y)$ is forced once the immoralities are determined.  Algorithms~\ref{MMPCstpred1},~\ref{MMPCstpred2} and~\ref{MMPCstpred3} consider the structures 3, 1 and 2 in figure~\ref{MMPCst2} in that order. Structure 3 indicates that if there is an immorality $X_1 - Y - X_2$, so that the directed edge set $D$ contains the edges $(X_1, Y) \in D$ and $(X_2, Y) \in D$ and undirected edges $\langle X_1 , W \rangle \in U$ and $\langle X_2, W\rangle \in U$, no edge between $X_1$ and $X_2$, then if there is an edge between $W$ and $Y$, it must be directed $(W,Y)$.

Since the third structure deals with directions forced directly through the presence of immoralities, it seems convenient to start with this one first; directing edges according to the first two structures will not produce additional immoralities, provided that there does exist a graph that is faithful to the distribution. 

Therefore, if appearances of the third structure are dealt with first, it will be unnecessary to consider the third structure again; no new examples will appear when additional edges are directed according to occurrences of the first and second structures. 

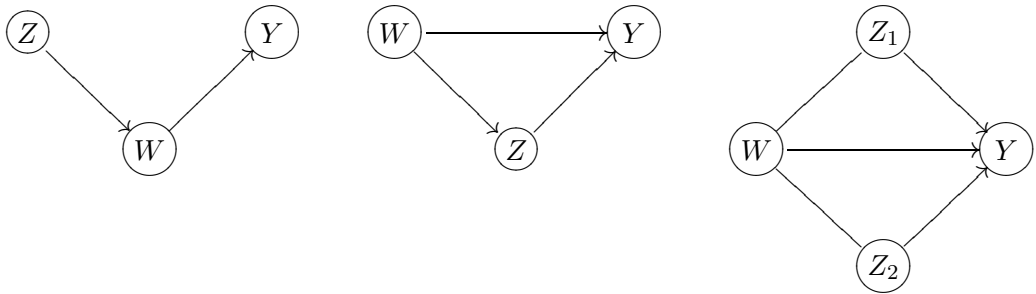
\begin{figure} 
\begin{center} 
\[ \UseTips \xymatrix{ 
*++[o][F]{Z}\ar[dr] &&*++[o][F]{Y} &   *++[o][F]{W}\ar[rr]\ar[dr] & & *++[o][F]{Y}  &
 &*++[o][F]{Z_1} \ar[dr] & \\
  & *++[o][F]{W}\ar[ur] & & 
&*++[o][F]{Z}\ar[ur] & &
*++[o][F]{W}\edge{ur}\edge{dr}\ar[rr]& & *++[o][F]{Y}\\ &&&&&&&*++[o][F]{Z_2}\ar[ur]& } \]
\end{center}
\caption{Strongly Protected Edges}\label{strongprotect} 
\end{figure}

It seems convenient to deal with the first structure of figure~\ref{strongprotect} secondly;  directing edges according to the second structure will not produce additional configurations that have the first structure. If  a direction $W \mapsto Y$ is directed in the second structure, thus playing the role of $Z \rightarrow W$ in the first structure, there is already an edge (which is $Z \rightarrow Y$ in the second structure) in the graph playing that role. 

Therefore, if the third structure is considered first and the first structure is considered second, it will not be necessary to search for these configurations again after considering the second structure. 

\paragraph{Stage 10} If the conditional dependence / independence statements produced are consistent with a probability distribution that has a faithful graphical representation, then the conditional independence statements are $d$-separation statements in a corresponding directed acyclic graph and hence no additional immoralities will be produced when the strongly protected edges are added in; the resulting graph will be the essential graph.

\begin{figure} 
\begin{center} 
\[ \UseTips \xymatrix{ 
*++[o][F]{X}\ar[r] & *++[o][F]{Y} \edge{r} &   *++[o][F]{Z}    & *++[o][F]{W}\ar[l]   } \]
\end{center} 
\caption{Example where additional immorality will be created}\label{figprob1} 
\end{figure}
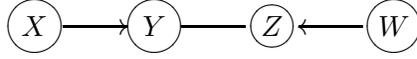

If they are not consistent, then it is possible that, using the procedure of the $M^3PC$ algorithm, additional immoralities will appear. This happens, for example, if a structure as in figure~\ref{figprob1} appears. If such a structure appears, then either $(X,Y,Z)$ or $(Y,Z,W)$ will appear as an immorality in the final graph.

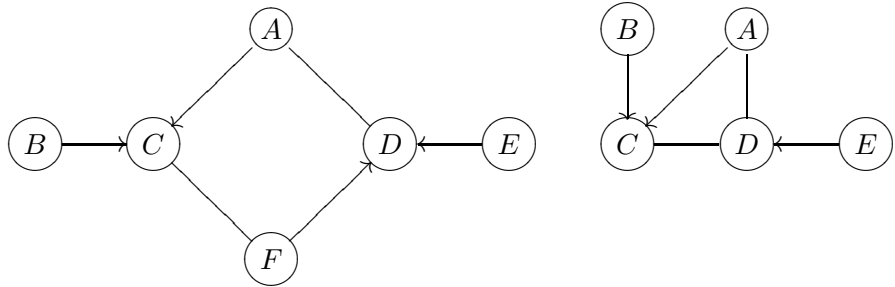
\begin{figure} 
\begin{center} 
\[ \UseTips \xymatrix{ && *++[o][F]{A} \ar[dl] \edge{dr}&&& *++[o][F]{B}\ar[d]& *++[o][F]{A}\ar[dl]&\\
*++[o][F]{B}\ar[r] & *++[o][F]{C} \edge{dr} &  & *++[o][F]{D}    & *++[o][F]{E}\ar[l] & *++[o][F]{C} & *++[o][F]{D}\edge{l}\edge{u} & *++[o][F]{E}\ar[l]\\ && *++[o][F]{F}\ar[ur] &&} \]
\end{center}
\caption{Examples where a directed cycle could be created by algorithm~\ref{MMPCstpred1}} \label{figprob2}  
\end{figure}

There are other problems that may appear if the set of dependence / independence relations does not correspond to a probability distribution with a faithful graphical model. The algorithm in stage 7 (algorithm~\ref{MMPCstpred1}) could produce cycles. This occurs if, for example, there is a structure of the type illustrated in figure~\ref{figprob2}.

If the graph returned after stage 9 satisfies the properties listed in theorem~\ref{thcheck}, and no additional immoralities have been produced in the direction of the edges, then stage 10 is unnecessary and the graph produced without stage 10 will be the essential graph of a Bayesian network which corresponds to the conditional independence / dependence relations that were derived from the data.

If the graph produced after stage 9 either contains directed cycles, or immoralities that were not in the original set of immoralities, or if the graph obtained by removing the directed edges is not triangulated, then step 10 is necessary. This step will produce a graph where the undirected edges may be oriented to produce the DAG of a Bayesian network where all the associations that were derived from data are present, but where  some of the conditional independence statements may not be represented by $d$-separation statements.

\begin{Th}\label{thcheck} 
\begin{enumerate}
\item The essential graph of a Bayesian network satisfies the three  conditions listed below.
\item If a graph with both directed and undirected edges, that has at most one edge between any pair of nodes satisfies the three properties listed below, then the undirected edges may be directed in such a way that the resulting directed graph is acyclic and does not contain additional immoralities. 
\end{enumerate}
The conditions are:   
\begin{enumerate}
\item The graph formed by removing the undirected edges does not contain a cycle.
\item The edges that appear in the structures shown in figure~\ref{strongprotect} are directed according to the directions given in figure~\ref{strongprotect} and
\item The connected components of the undirected graph obtained by removing the directed edges are triangulated.
\end{enumerate}
\end{Th}

\paragraph{Proof of theorem~\ref{thcheck}} {\bf Part 1}  All directed acyclic graphs that are Markov equivalent to each other have the same skeleton and the same immoralities (theorem~\ref{thmeq}). The essential graph of a Bayesian network is the graph where the directed edges are those that have that same direction in every directed acyclic graph that is Markov equivalent and all other edges are undirected (definition~\ref{defesgr}). The directed edges are the immoralities and the additional strongly protected edges. 

The first condition is satisfied, because otherwise all the Markov equivalent DAGs would contain a directed cycle, which is a contradiction. The second condition is satisfied, because these are the strongly protected edges, which have the same direction in every Markov equivalent DAG corresponding to the essential graph.

Suppose that the essential graph contains a cycle of undirected edges of length $\geq 4$ without a chord. Then in any corresponding DAG, the edges of the cycle must all be oriented the same way, otherwise there is at least one additional immorality. This implies that the DAG has a directed cycle, which is a contradiction.

Suppose that the essential graph contains a cycle of undirected edges of length $\geq 4$ with no undirected chords, but it contains  directed chords. Then the graph restricted to nodes of the cycle is triangulated, otherwise a structure of the first type in figure~\ref{strongprotect} appears with $W-Y$ undirected. Such structures do not appear.

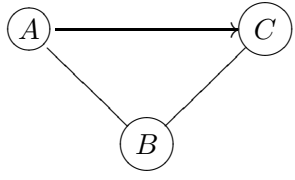
\begin{figure}
\begin{center} 
\[ \UseTips \xymatrix{   *++[o][F]{A} \ar[rr] \edge{dr} &&  *++[o][F]{C}\edge{dl}  \\
&*++[o][F]{B}& } \]
\end{center} 
\caption{Illustration for proof of theorem~\ref{thcheck}; $A \mapsto C$ is a directed chord in a cycle of undirected edges} \label{figill} 
\end{figure}

Therefore, there exists a structure of the type shown in figure~\ref{figill}. 

\noindent If the directed edge $A \mapsto C$ in the structure in figure~\ref{figill} appears as part of an immorality $A-C-D$ with a directed edge $D \mapsto C$ and no edge between $B$ and $D$, then $D-C-B$ is a structure of the first type in figure~\ref{strongprotect} and hence there is a contradiction, since $C \mapsto B$ would be strongly protected. If $A \mapsto C$ appears as part of an immorality $A-C-D$ with an undirected edge $B-D$, then the edge $B \mapsto C$ would be strongly protected, because it is part of a structure of the third type in figure~\ref{strongprotect}. Therefore $A \mapsto C$ is not part of an immorality in the essential graph.

If $A \mapsto C$ is not part of an immorality, but appears in the essential graph as part of a structure of the  first type, $E \mapsto A \mapsto C$, then there is an edge between $E$ and $B$; otherwise the edge $A \mapsto B$ would be strongly protected and hence directed, which is a contradiction. If $E-B$ is undirected, then the structure between $E$, $A$ and $B$ is of the type shown in figure~\ref{figill}. The direction $B \mapsto E$ is not permitted, since this would force $B \mapsto A$ as a strongly protected edge, which is a contradiction. If there is an edge $E \mapsto B$ in the essential graph, then there is a contradiction; either $B \mapsto C$ is strongly protected, or else there is an edge between $E$ and $C$ in the graph, contradicting the assertion that $E-A-C$ is a structure of the first type in figure~\ref{strongprotect}. It follows that $E-B$ is undirected and hence that $E-A-B$ forms a structure of the type in figure~\ref{figill} (two undirected edges, one directed edge). If $E \mapsto A$ is part of an immorality $E-A-F$, then there is an edge $F-B$, otherwise $A-B$ would be a strongly protected, and hence directed, edge. If $F \mapsto B$ is directed from $F$ to $B$, then $B \mapsto E$ is strongly protected and hence $B \mapsto A$ is strongly protected (structure 2 of figure~\ref{strongprotect}). If there is a direction $B \mapsto F$, then $B \mapsto A$ is strongly protected; $(A,B,F)$ form a structure 2 of figure~\ref{strongprotect}). Hence $F-B$ is undirected and $E,F$ play the roles of $Z_1, Z_2$ while $A,B$ play the roles of $Y,W$ in a structure 3 from figure~\ref{strongprotect} and hence $B \mapsto A$ is a strongly protected directed edge, which is a contradiction.

It follows that $E \mapsto A$ is not part of an immorality.

If $A \mapsto C$ is not part of an immorality, but appears in the essential graph as part of a structure of the second type, with a variable $F$ and directed edges $A \mapsto F$ and $F \mapsto C$, then there is an edge between $B$ and $F$. Otherwise $F \mapsto C - B$ is a structure of the first type and $C \mapsto B$ is strongly protected, which is a contradiction. $B-F$ is undirected; if it were directed $B \mapsto F$, then this would force $B \mapsto C$ as a strongly protected edge and if it were directed $F \mapsto B$, this would force $B \mapsto A$ as a strongly protected edge. Therefore the structure between $A,B$ and $F$ is of the type in figure~\ref{figill}.

By similar arguments as those above, $A \mapsto F$ does not appear as part of an immorality.

By induction, considering structures in the essential graph of the type in figure~\ref{figill}, if $A \mapsto C$ is not part of an immorality, then there is a similar adjacent structure, with a directed edge $E \mapsto A$ that is not part of an immorality. Working inductively, since there are no directed cycles, there is an infinite sequence of directed edges that are not part of an immorality, which is a contradiction.

It follows that an essential graph does not have a cycle of undirected edges of length $\geq 4$ either without a chord or with a directed chord. The connected components of the graph, formed by removing the directed edges, are therefore triangulated. 
  
\paragraph{Part 2} It is sufficient to show that the undirected edges in a graph satisfying the three conditions can be directed in a way that produces a directed acyclic graph, with no additional immoralities. 

Firstly, consider undirected edges that form part of a vee-structure. Take one pair, direct them as a chain connection $X \mapsto Y \mapsto Z$. If there is a vee-structure overlapping with one that has already been considered, direct the undirected edge in the structure in a way that does not produce an immorality.  If this procedure produces a cycle, then it follows that there is a cycle of undirected edges of length $\geq 4$ without a chord, which is a contradiction. 

Once the undirected edges that appear in vee - structures have been considered, any directions for the other edges that do not produce a cycle is valid. Such an ordering exists; consider a sequence where either orientation for an edge will produce a cycle, then there is already a directed cycle.

If the orientation of any of the edges produces an immorality with an edge that was directed in the original graph, then there is a vee -  structure $X \mapsto Y - Z$ with $X \mapsto Y$ and $Y-Z$ undirected, which is a contradiction.
 \qed 

\begin{Lmm}\label{lmmstage10} Stage 10 (algorithm~\ref{MMPCcheck}) produces the  essential graph of a Bayesian network, corresponding to a graphical model that contains all the conditional dependence statements that were derived earlier.
\end{Lmm}

\noindent {\bf Proof of lemma~\ref{lmmstage10}}   The addition of edges in the re-orientation stage ensures that the resulting graph will not contain conditional independence relations that are not in the data. Furthermore, these edges are necessary if a fork or chain connection is replaced with a collider. Arguments similar to those in the proof of theorem~\ref{thcheck} ensure that after directed cycles and additional immoralities have been dealt with, there is no directed edge that is a chord in an undirected cycle of length $\geq 4$. Furthermore, when edges are added during the triangulation procedure, arguments similar to those in the proof of theorem~\ref{thcheck} ensure that none of them appear in the position of $W-Y$ in the first or second structure of figure~\ref{strongprotect} and the additional edge $Z_1 - Z_2$ added when an undirected edge is added in the position of $W-Y$ in the structure of the third type in figure~\ref{strongprotect} during the triangulation ensures that no new strongly protected edges emerge during the triangulation. The resulting graph satisfies the criteria and, by theorem~\ref{thcheck}, is the essential graph of a Bayesian network.  \qed

\paragraph{Stage 7} This stage of the $M^3PC$ algorithm is listed as algorithm~\ref{MMPCstpred1}. All the immoralities have already been located in the previous stages; algorithm~\ref{MMPCstpred1} does an exhaustive search for configurations where the direction of an edge $W \mapsto Y$ is forced corresponding to the third structure of figure~\ref{strongprotect}.  For each immorality $(X_i, X_j, X_k) \in {\cal I}$, enumerating the set ${\cal Z}_i \cap {\cal Z}_j \cap {\cal Z}_k$ as 
\[ {\cal Z}_i \cap {\cal Z}_j \cap {\cal Z}_k = \{X_{\alpha(1)}, \ldots, X_{\alpha(p)}\},\] 

\noindent the directed edges $\{(X_{\alpha(1)}, X_j), \ldots, (X_{\alpha(p)}, X_j)\}$ are added to $D$ and  the undirected edges \[ \{\langle X_{\alpha(1)}, X_j\rangle, \ldots, \langle X_{\alpha(p)}, X_j \rangle\}\] 

\noindent removed from $U$.  

\paragraph{Stage 8} This stage of the $M^3PC$ algorithm is listed as  algorithm~\ref{MMPCstpred2}.  
It does an exhaustive search for structures of the   first type in  figure~\ref{strongprotect}. The following loop is carried out until no more structures of the type are located: for each directed edge $(X_i, X_j) \in D$, with the set  ${\cal Z}_i \backslash   {\cal Z}_j$ enumerated as ${\cal Z}_i \backslash   {\cal Z}_j = \{X_{\alpha(1)}, \ldots, X_{\alpha(k)}\}$, if $\langle X_j, X_{\alpha(l)} \rangle \in U$, then it is removed from $U$ and the directed edge $(X_j, X_{\alpha(l)})$ added to $D$.   

When considering the first structure, directing an edge $W \mapsto Y$ will not create any additional immoralities if there is a faithful graph, since all the immoralities have already been located. If a new immorality is produced by the directing of an edge at this stage, it means that there is not a faithful graph structure to represent the set of conditional independence statements.
	
After this stage, there are no structures on three nodes $W, Z, Y$ with directed edge $(Z, W)$, undirected edge $\langle W, Y\rangle$ and no edge between $Z$ and $Y$.

\paragraph{Stage 9} This stage of the $M^3PC$ algorithm is listed as algorithm~\ref{MMPCstpred3}. It deals exhaustively with structures of the second type in figure~\ref{strongprotect}. For each directed edge $(X_i, X_j) \in D$, let 
\[ {\cal Y}_j = \{ X_k | (X_j, X_k) \in D\},\] the set of directed edges from $X_j$ to $X_k$.   Then, for each $X_j \in {\cal Y}_j \cap {\cal Z}_i$,  the edge $(X_i, X_k)$ is added to $D$ and the edge $\langle X_i, X_k \rangle$ is removed from $U$, the set of undirected edges. This is repeated until there is a cycle where no additional edges are directed.

After this stage, there are no longer any triples of nodes $W, Z, Y$ with directed edges $(W, Z)$ and $(Z , Y)$ and an undirected edge $\langle W , Y \rangle$. If there is a faithful graph, then this stage does not introduce new immoralities and hence no further strongly protected edges of the type shown in the third structure.

Furthermore, directing the edge $W \rightarrow Y$ in the second structure cannot create additional structures between three nodes $X_1,X_2,X_3$ with a directed edge $(X_1,X_2)$ and an undirected edge $\langle X_2,X_3\rangle$ and no edge between $X_1$ and $X_3$;  the edge $(X_1, X_2)$ has just been directed  and, if it has been directed according to the principle in the second structure, there is already a node $X_4$ forming a directed edge $(X_4, X_2)$ (where $X_4$ plays the role of $Z$ in the second structure). Therefore the edge between $X_2$ and $X_3$ has already been directed in the previous stage. 

\begin{algorithm}
\caption{The Modified MMPC Algorithm: determining $A(i,j;S)$}
\label{MMPCmodalg0}
\begin{algorithmic}
\STATE initialisation $A(i,j;k_1, \ldots, k_m) = 2$ for all $m \geq 0$ and all $(i,j,k_1,\ldots, k_m) \in \{1,\ldots, d\}^{m+2}$
\STATE $A(i,j;S) = 0$ represents $X_i \perp X_j | S$; $A(i,j;S) = 1$ represents $X_i \not \perp X_j | S$
\FOR{$i = 0,\ldots, m$}
\FOR{each permutation $\sigma$ of $k_1,\ldots, k_{m+2}$ of  length $i+2$}
\IF{$A(k_{\sigma(1)}, k_{\sigma(2)}; k_{\sigma(3)}, \ldots, k_{\sigma(i+2)}) = 0$ or $1$}
\STATE do nothing
\ELSE 
\IF{Condition $\Xi(k_{\sigma(1)},k_{\sigma(2)};k_{\sigma(3)}, \ldots, k_{\sigma(i+2)})$ (definition~\ref{defcondxi}) holds} 
\STATE Set $B(k_{\sigma(1)},k_{\sigma(2)}; k_{\sigma(3)}, \ldots, k_{\sigma(m+2)})   = 0$
\ELSE
\STATE Set $B(k_{\sigma(1)},k_{\sigma(2)}; k_{\sigma(3)}, \ldots, k_{\sigma(m+2)})  = 1$
\ENDIF
\ENDIF
\ENDFOR
\FOR{each permutation $\sigma$ of $k_1,\ldots, k_{m+2}$ of  length $i+2$}
\IF{$B(k_{\sigma(1)},k_{\sigma(2)}; k_{\sigma(3)}, \ldots, k_{\sigma(m+2)}) = 1$}
\STATE Set  $A(k_{\sigma(1)},k_{\sigma(2)}; k_{\sigma(3)}, \ldots, k_{\sigma(m+2)}) = 1$ and $A(k_{\sigma(2)},k_{\sigma(1)}; k_{\sigma(3)}, \ldots, k_{\sigma(m+2)}) = 1$
\ELSE 
\FOR{$j=3, \ldots, m$}
\IF{$B(k_{\sigma(1)},k_{\sigma(j)}; \{k_{\sigma(2)}, \ldots, k_{\sigma(m+1)}\}\backslash \{k_{\sigma(j)}\}) = 0$ and\\ ($A(k_{\sigma(1)},k_{\sigma(2)}; \phi) = 1$ or $A(k_{\sigma(1)}, k_{\sigma(j)};\phi) = 1$)}
\STATE Set $B(k_{\sigma(1)},k_{\sigma(2)}; k_{\sigma(3)}, \ldots, k_{\sigma(m+2)}) = 1$
\ENDIF
\IF{$B(k_{\sigma(2)},k_{\sigma(j)}; \{k_{\sigma(1)},k_{\sigma(3)}, \ldots, k_{\sigma(m+1)}\}\backslash \{k_{\sigma(j)}\}) = 0$ and\\ ($A(k_{\sigma(1)},k_{\sigma(2)}) = 1$ or $A(k_{\sigma(2)}, k_{\sigma(j)}) = 1$)}
\STATE Set $B(k_{\sigma(1)},k_{\sigma(2)}; k_{\sigma(3)}, \ldots, k_{\sigma(m+2)})  = 1$
\ENDIF
\ENDFOR
\STATE Let $A(k_{\sigma(1)},k_{\sigma(2)}; k_{\sigma(3)}, \ldots, k_{\sigma(m+2)}) =  B(k_{\sigma(1)},k_{\sigma(2)}; k_{\sigma(3)}, \ldots, k_{\sigma(m+2)})$
\STATE and $A(k_{\sigma(2)},k_{\sigma(1)}; k_{\sigma(3)}, \ldots, k_{\sigma(m+2)}) =B(k_{\sigma(1)},k_{\sigma(2)}; k_{\sigma(3)}, \ldots, k_{\sigma(m+2)})$
\ENDIF
\ENDFOR
\ENDFOR
 \end{algorithmic}
\end{algorithm}

\begin{algorithm}
\caption{The Modified MMPC Algorithm, Stage 1}
\label{MMPCmodalg1}
\begin{algorithmic}
\STATE For a subset $S = \{X_{k_1}, \ldots, X_{k_m}\} \subset \{X_1, \ldots, X_d\}$, let $A(i,j;S)$ and $A(i,j;k_1,\ldots, k_m)$ both denote the same thing. $A(i,j;S) = 0$ if $X_i \perp X_j | \{X_{k_1}, \ldots, X_{k_m}\}$ is included in the set of dependencies and $A(i,j;S) = 1$ if $X_i \perp X_j | \{X_{k_1}, \ldots, X_{k_m}\}$ is rejected from the set of dependencies.
\FOR{$j = 1, \ldots, d$ } 
 \STATE Set ${\cal Z}_{j,0} = \phi$ (the empty set)
 \FOR{$i = 1,\ldots, d$}
 \IF{$i = j$} \STATE Set ${\cal Z}_{j,i} = {\cal Z}_{j,i-1}$
 \ELSE
 \IF {$A(i,j; {\cal Z}_{j, i-1}) = 0$  } \STATE Set ${\cal Z}_{j,i} = {\cal Z}_{j,i-1}$, Set $S_{ji} = S_{ij} = {\cal Z}_{j,i}$
\ELSE \STATE Set ${\cal Z}_{j,i} = {\cal Z}_{j, i-1} \cup \{X_i\}$
  \ENDIF
  \ENDIF
 \ENDFOR
\STATE Set ${\cal Z}_j = {\cal Z}_{j,d}$.
\ENDFOR  
\STATE {\bf Note}: $S_{ij}$ is the set such that $X_i \perp X_j | S_{ij}$ that determines that the edge $\langle X_i, X_j \rangle$ is not in the skeleton
\end{algorithmic}
\end{algorithm}

\begin{algorithm}
\caption{The Modified MMPC Algorithm, Stage 2}
\label{MMPCmodalg2}
\begin{algorithmic}
\STATE Recall $A(i,j;S) = 0$ if $X_i \perp X_j | S$ has been accepted; $A(i,j;S) = 1$ if $X_i \not \perp X_j | S$ has been accepted.
\STATE Set ${\cal I} = \phi$.  By the end of stage 5 (algorithm~\ref{MMPCmodalg5}) of the $M^3PC$ algorithm, this will be the set of immoralities.
\FOR{$j = 1,\ldots, d$}  \STATE Set ${\cal Y}_{j,0} = {\cal Z}_{j}$ 
\FOR{$k = 1,\ldots, d$}
\IF{$X_j \not \in {\cal Z}_{k}$}
\FOR{$i = 1,\ldots, d$}
\IF{$X_i \in {\cal Z}_{j} \cap {\cal Z}_{k}$}
\IF{$X_i \not \in {\cal Z}_{k,j-1}$  }
\STATE Let ${\cal I}  = {\cal I}  \cup \{(X_j, X_i, X_k)\}$ 
\ELSE \STATE Leave ${\cal I}$ unaltered. 
\ENDIF
\ENDIF
\ENDFOR 
\STATE Set ${\cal Y}_{j,k} = {\cal Y}_{j,k-1} \backslash \{X_k\}$
\ELSE \STATE Set ${\cal Y}_{j,k} = {\cal Y}_{j,k-1}$
\ENDIF
\ENDFOR
\ENDFOR
\STATE Set ${\cal Z}_j = {\cal Y}_{j,d}$
\end{algorithmic}
\end{algorithm}

\begin{algorithm}
\caption{The Modified MMPC Algorithm, Stage 3}
\label{MMPCmodalg3}
\begin{algorithmic}
\STATE Recall $A(i,j;S) = 0$ if $X_i \perp X_j | S$ has been accepted; $A(i,j;S) = 1$ if $X_i \perp X_j | S$ has been rejected.
\FOR{$j = 1,\ldots, d$}  \STATE Set ${\cal Y}_{j,0} = {\cal Z}_j$ 
\FOR{$i = 1,\ldots, d$}
\IF{$X_i \in {\cal Y}_{j,i-1}$ and there exists a subset $S \subseteq {\cal Y}_{j,i-1} \backslash \{X_i\}$ such that $A(i,j;S) = 0$}
\STATE   Set 
${\cal Y}_{j, i} = {\cal Y}_{j,i-1} \backslash \{X_i\}$ 
\STATE Let $S_{ij} = S$.
\FOR{$p = 1, \ldots, d$}
\IF{$X_p \in ({\cal Y}_{j, i}\backslash S) \cap {\cal Z}_i$ and $X_p \not \in S$ }
\STATE Let ${\cal I}  = {\cal I}  \cup \{(X_j, X_p, X_i)\}$ 
\ELSE \STATE leave ${\cal I}$ unaltered.
\ENDIF
\ENDFOR
\ELSE  
  \STATE Set ${\cal Y}_{j,i} = {\cal Y}_{j,i-1}$
\ENDIF
\ENDFOR
\ENDFOR
\STATE Set ${\cal Z}_j = {\cal Y}_{j,d}$
\end{algorithmic}
\end{algorithm}

\begin{algorithm}
\caption{The Modified MMPC Algorithm, Stage 4}
\label{MMPCmodalg4}
\begin{algorithmic}
\FOR{$j = 1,\ldots, d$} \STATE Set ${\cal Y}_{j,0} = {\cal Z}_j$.
\FOR{$k = 1,\ldots, d$}
\IF{$X_k \in {\cal Y}_{j,k-1}$ but $X_j \not \in {\cal Z}_k$}
\STATE Set ${\cal Y}_{j,k} = {\cal Y}_{j,k-1} \backslash \{X_k\}$
\ELSE \STATE Set ${\cal Y}_{j,k} = {\cal Y}_{j,k-1}$
\ENDIF
\ENDFOR
\STATE Set ${\cal Z}_j = {\cal Y}_{j,d}$
\ENDFOR
\end{algorithmic}
\end{algorithm}

\begin{algorithm}
\caption{The Modified MMPC Algorithm, Stage 5}
\label{MMPCmodalg5}
\begin{algorithmic}
\STATE This algorithm removes those elements from ${\cal I}$ that are not immoralities in the final graph and ensures that an immorality appears in ${\cal I}$ listed as $(X_i,Y,X_j)$ for $i < j$. 
\FOR{$i = 1, \ldots, d$}
\FOR{$(j,k) \in \{1, \ldots, d\}^2, i \neq j, i \neq k, j \neq k$}
\IF{$(X_j, X_i, X_k) \in {\cal I}$ and $X_i \not \in {\cal Z}_j \cap {\cal Z}_k$}
\STATE Let ${\cal I} = {\cal I} \backslash \{(X_j, X_i, X_k)\}$
\ENDIF
\ENDFOR
\ENDFOR
\FOR{$i = 1, \ldots, d$}
\FOR{$1 \leq j < k \leq d$}
\IF{$(X_k, X_i, X_j) \in {\cal I}$}
\STATE Let ${\cal I} = ({\cal I} \cup \{(X_j, X_i, X_k)\})\backslash \{(X_k, X_i, X_j)\}$
\ENDIF
\ENDFOR
\ENDFOR
\STATE The next stage is inserted in case the set of $A(i,j;S)$ that have been established do not correspond to a distribution that has a faithful graphical model, in which case some of the immoralities may be incompatible.
\FOR{$1 \leq i < k \leq d$}
\FOR{$j=1, \ldots, d$; $j \neq i$, $j \neq k$}
\IF{$(X_i, X_j, X_k) \in {\cal I}$}
\FOR{$p = 1,\ldots, d$}
\IF{$(X_j, X_k, X_p) \in {\cal I}$ or $(X_p, X_k, X_j) \in {\cal I}$}
\IF{$A(j,p; S_{jp} \cup \{X_k\}) = 0$}
\STATE Set ${\cal I} = {\cal I} \backslash \{(X_j, X_k, X_p), (X_p, X_k, X_j)\}$
\ELSE \STATE Set ${\cal I} = {\cal I} \backslash \{(X_i, X_j, X_k)\}$
\ENDIF
\ENDIF
\ENDFOR
\ENDIF
\ENDFOR
\ENDFOR
\end{algorithmic}
\end{algorithm}

 \begin{algorithm}
\caption{The Modified MMPC Algorithm, Stage 6}
\label{MMPCmodalg6}
\begin{algorithmic}
\STATE Set ${\cal U}_1 = {\cal Z}_1$
\FOR{$i = 2,\ldots,d$} 
\STATE Set ${\cal U}_i =  {\cal Z}_i \backslash \{X_1,\ldots, X_{i-1}\}$. 
\ENDFOR
\STATE By the end of the algorithm, the essential graph will contain an undirected edge $\langle X_i, X_j \rangle$ if and only if $X_j \in {\cal U}_i$ for $j > i$ and $X_j \not \in {\cal U}_i$ for $j < i$. 
\STATE Set ${\cal P}_1 = \ldots = {\cal P}_d = \phi$. By the end of the graph, these will be the sets of variables such that, in the essential graph, there is a directed edge $Y \mapsto X_i$ if and only if $Y \in {\cal P}_i$. 
\FOR{$i = 1, \ldots, d-1$}
\FOR{$k = i+1, \ldots,d$}
\FOR{$j = 1, \ldots, d$}
\IF{$(X_i, X_j, X_k) \in {\cal I}$}
\STATE Let ${\cal U}_i = {\cal U}_i \backslash \{X_j\}$, let ${\cal U}_k = {\cal U}_k \backslash \{X_j\}$, let ${\cal U}_j = {\cal U}_j \backslash \{X_i, X_k\}$. 

Let ${\cal P}_j = {\cal P}_j \cup \{X_i, X_k\}$
\ENDIF
\ENDFOR
\ENDFOR
\ENDFOR
\end{algorithmic}
\end{algorithm}

\begin{algorithm}
\caption{Locating Additional Strongly Protected Edges, Stage 1}
\label{MMPCstpred1}
\begin{algorithmic}
\FOR{$i = 1,\ldots, d-1$}
\FOR{$k = i+1, \ldots, d$}
\FOR{$j = 1, \ldots, d$}
\FOR{$l = 1,\ldots, d$}
\IF{$X_l \in {\cal Z}_i \cap {\cal Z}_j \cap {\cal Z}_k$}
\STATE Let ${\cal P}_j = {\cal P}_j \cup \{ X_l \}$, ${\cal U}_l = {\cal U}_l \backslash \{X_j\}$ and ${\cal U}_j = {\cal U}_j \backslash \{   X_l  \}$
\ENDIF
\ENDFOR
\ENDFOR
\ENDFOR
\ENDFOR
\end{algorithmic}
\end{algorithm}

\begin{algorithm}
\caption{Locating Additional Strongly Protected Edges, Stage 2}
\label{MMPCstpred2}
\begin{algorithmic}
\STATE Let $D_2 = 1$ 
\REPEAT
\STATE Let $D_1 = 0$
\FOR{$i = 1, \ldots, d$}
\FOR{$j = 1, \ldots, d$}
\IF{$(X_i, X_j) \in D$}
\FOR{$k = 1,\ldots, d$}
\IF{$X_k \in {\cal Z}_j \backslash {\cal Z}_i$}
\STATE Set ${\cal P}_k = {\cal P}_k \cup \{X_j\}$,   $D_1 = D_1 + 1$, ${\cal U}_j = {\cal U}_j \backslash \{X_k\}$, ${\cal U}_k = {\cal U}_k \backslash \{X_j\}$ 
\ENDIF
\ENDFOR
\ENDIF
\ENDFOR
\ENDFOR
\STATE Let $D_2 = D_1$
\UNTIL{$D_2 = 0$}
\end{algorithmic}
\end{algorithm}

\begin{algorithm}
\caption{Locating Additional Strongly Protected Edges, Stage 3}
\label{MMPCstpred3}
\begin{algorithmic}
\STATE Let $D_2 = 1$ 
\REPEAT
\STATE Let $D_1 = 0$
\FOR{$i= 1, \ldots, d$}
\FOR{$j = 1,\ldots, d$}
\IF{$(X_i, X_j) \in d$}
\FOR{$k = 1,\ldots, d$}
\IF{$(X_j, X_k) \in D$ and $X_j \in {\cal Z}_i$}
\STATE Let ${\cal P}_k = {\cal P}_k \cup \{ X_i \}$, $D_1 = D_1 + 1$, let ${\cal U}_i = {\cal U}_i \backslash \{ X_k   \}$, let ${\cal U}_k = {\cal U}_k \backslash \{X_i\}$.
\ENDIF
\ENDFOR
\ENDIF
\ENDFOR
\ENDFOR
\STATE Let $D_2 = D_1$
\UNTIL{$D_2 = \phi$}
\end{algorithmic}
\end{algorithm}
 
\begin{algorithm}
\caption{Locating Problems Caused by non-existence of Faithful graphical model}
\label{MMPCcheck}
\begin{algorithmic}
\STATE Recall that ${\cal I}$ is the set of immoralities obtained from data. Let ${\cal D}_0$ denote the set of edges such that $(X,Y) \in {\cal D}_0$ if and only if either $(X,Y,Z) \in {\cal I}$ or $(Z,Y,X)  \in {\cal I}$ for some $Z$. Let ${\cal E}$ denote the set of edges in the skeleton before this stage of the algorithm.
\REPEAT
\REPEAT \STATE {\bf Locate} a cycle of directed edges. Select at random an edge in the cycle that does not belong to ${\cal D}_0$. Change the orientation of the edge. If this results in an immorality $(X,Y,Z)$, then add a directed edge $X \mapsto Z$ or $Z \mapsto X$, choosing a direction, if possible, that does not create a cycle.  
\UNTIL there are no directed cycles.
\REPEAT \STATE Locate ${\cal K}$, the set of immoralities $(X,Y,Z)$ such that $(X,Y,Z) \not \in {\cal I}$, and $\{\langle X,Y \rangle, \langle Y,Z \rangle \} \subset {\cal E}$. 
   \STATE for each immorality  $(X,Y,Z) \in {\cal K}$ 
add either the directed edge $X \mapsto Z$ or $Z \mapsto X$ in such a way that no directed cycles are created. (This is clearly possible - if both directions cause a directed cycle, then there is already a directed cycle present).
\UNTIL all the immoralities in ${\cal K}$ have been dealt with.
\STATE Perform algorithms~\ref{MMPCstpred1},~\ref{MMPCstpred2},~\ref{MMPCstpred3}, locating and orienting undirected edges that take the position $(W,Y)$ in one of the three structures in figure~\ref{strongprotect} and directing them $W \mapsto Y$ according to figure~\ref{strongprotect}. 
\UNTIL there is a full run in which no changes are made.
\STATE {\bf Add in the necessary undirected edges} to triangulate the sub-graph obtained by removing the directed edges, adding in the undirected edge $Z_1-Z_2$ if this involves adding in an undirected edge in the position taken by $W-Y$ in structure 3 of figure~\ref{strongprotect}.  
\end{algorithmic}
\end{algorithm}

\section{Time Complexity of the Algorithms} 
A common measure of the `complexity' (or time required) for an algorithm is the number of {\em statistical calls} that are made (definition~\ref{defstatcall}). 
This is discussed in Spirtes, Glymour and Scheines~\cite{SGS} and Margaritas and Thrun~\cite{MT}. Both of these use the $G^2$ statistic (definition~\ref{defteststat} equation (\ref{eqtestgee})), calculating the $p$ value and accepting independence if the $p$ value is less than the nominal level $\alpha$, usually $\alpha = 0.05$. 

If the number of variables is very large and the number of instantiations is very large, then a more accurate picture of the complexity would be the number of statistical calls necessary, where a statistical call is defined as a reference to the data set in order to compute the marginal cell counts for a subset of the variables.

Recall that the $M^3$PC algorithm  locates the skeleton and immoralities through algorithms~\ref{MMPCmodalg1}, ~\ref{MMPCmodalg2}, ~\ref{MMPCmodalg3}, ~\ref{MMPCmodalg4}, ~\ref{MMPCmodalg5} and~\ref{MMPCmodalg6}. Of these, algorithms~\ref{MMPCmodalg5} and~\ref{MMPCmodalg6} are purely algebraic rearrangements, checking whether or not the edges associated with immoralities obtained previously have been deleted and if not putting the edges into the directed edge sets. 
Stage 4, given by algorithm~\ref{MMPCmodalg4} is also purely algebraic, removing $X_j$ from ${\cal Z}_k$ if $X_k \not \in {\cal Z}_j$. These parts of the algorithm do not have significant computational  complexity; they do not require tests of conditional independence. Following the criteria of Spirtes, Glymour and Scheines, and also Margaritas and Thrun, the computation complexity of the algorithm comes from the conditional independence tests and these are all located in stages 1,2 and 3;  algorithms~\ref{MMPCmodalg1},~\ref{MMPCmodalg2} and~\ref{MMPCmodalg3} respectively.  

Stages, stages 7, 8 and 9 (finding the additional strongly protected edges) given by algorithms~\ref{MMPCstpred1}, ~\ref{MMPCstpred2} and~\ref{MMPCstpred3} respectively are algebraic, locating those edges whose directions are logically forced by the immoralities. 

If the set of conditional independence statements does not correspond to a distribution with a faithful graphical representation, then stage 5 (algorithm~\ref{MMPCmodalg5}) may require some additional tests; the number will increase depending on the level of inaccuracy of the faithfulness assumption.

If the complexity is measured by the number of test statistics (definition~\ref{defteststat}), then the comparison is as follows. 

\paragraph{Stage 1} (algorithm~\ref{MMPCmodalg1}) If the test procedure outlined in~\cite{TBA} is used, there is an upper bound of  $d(d-1)$ conditional independence tests. This may be reduced if it turns out that a test for conditional independence has already been made. This is the same as the first stage of the MMPC algorithm. 

\paragraph{Stage 2} (algorithm~\ref{MMPCmodalg2}) does not require additional statistical calls. For the current neighbour sets ${\cal Z}_1, \ldots {\cal Z}_d$, it considers situations where $X_j \not \in {\cal Z}_k$ and removes the variable $X_k$   from ${\cal Z}_j$ if $X_k \in {\cal Z}_j$. If a variable $X_i \in {\cal Z}_j \cap {\cal Z}_k$, it has to be determined whether or not the vee - structure (definition~\ref{defveestr}) $X_j - X_i - X_k$ is  an immorality. This is done by determining whether or not $X_i$ is in a set that has already been determined in stage 1.\vspace{5mm}

\noindent For determining upper bounds on the number of statistical calls required in the next stages, the following definition

\begin{Defn}[Maximal size of neighbour sets]\label{defceetheta} Let $C = \max_{j=1, \ldots, d}|{\cal Z}_j|$, denote the size of the largest set obtained after stage 1 of the $M^3PC$ algorithm.
 
Let $\Theta$ denote the size of the largest parent / child set in the essential graph of a Bayesian network.

Let $P$ denote the size of the largest parent set for a variable among all the faithful DAGs. 
\end{Defn}

\paragraph{Stage 3} (algorithm~\ref{MMPCmodalg3}) This is the computationally most expensive part of the algorithm. As pointed out in section 7 of~\cite{TBA}, the complexity for constructing the skeleton (that is, deciding which nodes to remove from ${\cal Z}_j$) is $O(2^{|{\cal Z}_j|})$, where $|{\cal Z}_j|$ is the number of variables in ${\cal Z}_j$. Working through the variables in order, a variable $X_i$ is removed from ${\cal Z}_j$ if there exists a subset   $S \subseteq {\cal Z}_j \backslash \{X_i\}$ (where ${\cal Z}_j$ represents the current set of variables that have not yet been eliminated) such that $X_j \perp X_i | S$. This involves working through the subsets until one is found, or working through all the subsets if there is an edge $\langle X_i, X_j \rangle$ in the skeleton. There are $2^{|{\cal Z}_j|}$ possible subsets of ${\cal Z}_j$. The article~\cite{TBA} therefore states the bound as $O(d 2^C)$, where $C$ is the upper bound in definition~\ref{defceetheta}. If $d$ is large, then  the algorithm is clearly not feasible if $C$ is of order $d$. 

Furthermore, there is a limit, depending on the size of the data set, on the size of the conditioning set for which independence tests can be made.  For a test of $H_0: X \perp Y|S$, it is required that for all $(x,y)$, $N_{X,Y,S}(x,y;\underline{s}) \geq 5$ (where $N_{X,Y,S}(x,y;\underline{s})$ is defined in definition~\ref{defteststat}) for at least one instantiation $\underline{s}$ of $S$. This is the criterion suggested and applied in~\cite{TBA}. It is necessary for the $\chi^2$ distribution to have any level of accuracy, but for a fixed number of data, the larger the number of states that the conditioning set $S$ can take, the lower the {\em power} of the test and the smaller the probability that $H_0$ is true when the result of the test is `insufficient information to reject $H_0$'.  

An upper bound of size $l$, for a fixed $l$ that is determined either by the size of the data set, or else computational cost or both, has to be placed on the subsets taken from ${\cal Z}_j$ for each $j = 1, \ldots, d$. In this case, there is an upper bound on the number of subsets to be considered given by
\[ d\times \sum_{k=0}^l \left(\begin{array}{c} C \\ k \end{array}\right) \leq deC^l.\]

\noindent This is in line with the order stated in~\cite{TBA}. 

In practise, it is often observed that even though the parents and children set of $X_j$ is  a subset of ${\cal Z}_j$ obtained after stage 2  nevertheless,  $\max_{j=1, \ldots, d}|{\cal Z}_j|$ (the maximum size of ${\cal Z}_j$, $j = 1, \ldots, d$ obtained after stage 2) is often of the same order as the size of the largest parents / children set, $\Theta$ (definition~\ref{defceetheta}). That is, although $\Theta \leq C$ it usually turns out that $C = O(\Theta)$. This remark is made in~\cite{TBA} section 7.

In line with~\cite{TBA}, the complexity of the algorithm for locating the skeleton is bounded above by
\begin{equation}\label{expmmpc}  d(d-1) + edC^l
\end{equation}

\noindent where $l$ is the size of the largest conditioning set permitted, $d$ is the number of variables and $C$ is the number of variables in the largest set ${\cal Z}_j$ obtained after stage 2. 

The additional complexity required to obtain the immoralities is small by comparison; no additional statistical calls are required. 

It follows that the number of conditional independence tests required for the $MMPC$ algorithm and for the $M^3PC$ algorithm  when the modification for incorporating immoralities is included, but the modification to the independence tests is not included, is bounded above by  expression (\ref{expmmpc}).  Using this measure of complexity, both algorithms have equal complexity, but the modification requires additional algebraic stages.

\paragraph{Complexity with the Modified Procedure to Establish the Conditional Independence Statements}

The modification that alters the way that conditional independence is determined, yields a substantial increase in the complexity. The example in subsection~\ref{subsWAM} illustrates, on an example with $6$ variables with $1000$ observations, that running the algorithm without the modification can lead to the removal of edges that result in a $d$-separation statement that is present in the graphical model,  when the corresponding independence statement has been rejected at the $5\%$ significance level. The measure required to circumvent this measure is computationally expensive; in situations where adequate results are achieved without the modification, the additional accuracy may not be justified by the increase in complexity.  

 If $X_i \perp X_j | \{X_{k_1}, \ldots, X_{k_m}\}$ is considered, there are
\[  \sum_{j=0}^m \left(\begin{array}{c} m \\ j \end{array}\right) \left(\begin{array}{c} j+2 \\ 2 \end{array}\right) \leq (m+2)(m+1)2^{m-1}\]
\noindent tests to be carried out. Let $C = \max\{|{\cal Z}_1|, \ldots, |{\cal Z}_d|\}$ be the maximum size of the sets obtained after the first stage of the algorithm, then it follows that there is an upper bound of
\begin{equation}\label{exmodtest} d(d-1)(C+2)(C+1)2^{C-1}\end{equation}

\noindent This corresponds to an increase in the computational complexity by a factor of 
\begin{equation}\label{exccfac} (C+2)(C+1)2^{C-1}.
 \end{equation}

\noindent on the number of tests required to perform the first stage of the algorithm if no limit is put on the size of the conditioning sets. This upper bound is substantially larger than the true number of statistical calls required; it does not take into account that ${\cal Z}_i \cap {\cal Z}_j \neq \phi$ for some $(i,j)$ and that once the value of  $A(i,j;S)$ is registered, it is not computed again. \vspace{5mm}

\noindent If a limit of size $l$ is put on the size of the conditioning sets,  then for a neighbour set of size $C$ where $C \geq l$, there are $\left(\begin{array}{c} C \\ l+1 \end{array}\right)$ values of $A(p,q;S)$ to be established to determine whether or not $X_k \in {\cal Z}_j$ if the modification to the independence testing is employed.  There is therefore an upper bound of 

\begin{equation}\label{equppbd}  d(d-1)\left(\begin{array}{c} C \\ l+1 \end{array}\right)(l+2)(l+1)2^{l-1} \leq  d(d-1)\frac{C^l}{l!}(l+2)2^{l-1} \end{equation}

\noindent on the number of chi squared tests required to complete stage 1 of the $M^3PC$ algorithm. After these tests are performed, it is clear that no further conditional independence tests are required. 

If a limit $l$ is put on the size of conditioning sets, then if $|{\cal Z}_j| >  l$, it follows that an upper bound of $\left(\begin{array}{c}|{\cal Z}_j| \\ l \end{array}\right)$ tests   are required in stage $1$ to determine that a variable should be added to the neighbour set. There is clearly an upper bound of 

\begin{equation}\label{eqpropuppbd}
 \frac{C^l}{l!}(l+2)2^{l-1}
\end{equation}

\noindent on the factor by which the computational complexity is increased when the   modification for determining conditional independence is adopted. Note that when the   modification for determining conditional independence is applied,  all the conditional independence statements required are computed in stage 1 of the algorithm.

\paragraph{Alternative Complexity Measures} If the complexity is measured by the number of statistical calls, where a statistical call is defined as a computation of the marginal empirical probability distribution over a set of variables that requires reference to the data set, then all the statistical calls are made in stage 1 of the algorithm; there is an upper bound of $d(d-1)$. There is no difference in the number of statistical calls required for the MMPC and the modified MMPC. The modification to the method for testing conditional independence does not require any further calls to the data set; if it is required to determine a conditional independence relation for a set of variables ${\cal X}$, then the empirical probability distribution for any subset of ${\cal X}$ can be derived from the empirical distribution over ${\cal X}$. 

It is expected that, as the number of variables and number of instantiations increases, the determining factor will increasingly be the number of calls to the data set.

\section{Related Work}  
The work in this article is motivated by the MMPC part of the MMHC algorithm of~\cite{TBA}, noting that with a relatively small addition in the computational complexity, the essential graph can be recovered by the  modification proposed  for the $M^3PC$ algorithm that locates the immoralities, thus rendering the edge orientation phase (the most time consuming phase) of the  Maximum Minimum Hill Climbing algorithm of~\cite{TBA} unnecessary. This modification leads to a substantial improvement in the time taken. 
The modification introduced to ensure that the conditional independence statements are consistent with the dependence statements derived from data improves the accuracy of the algorithm, as illustrated in the example in subsection~\ref{subsWAM}. 

As mentioned in section~\ref{secintro}, the main approaches to the problem of locating the graph structure are constraint based (testing for conditional independence), search-and-score (maximising an appropriate score function) and hybrid (methods that use both constraint based ideas and search and score ideas). The Maximum Minimum Hill Climbing algorithm presented in~\cite{TBA} is a hybrid algorithm, using the constraint based MMPC procedure to obtain the skeleton and, once the sets of neighbours for each variable is established,  it uses a greedy search and score algorithm, starting with an empty graph and, using the Bayesian Dirichlet metric (discussed later) for scoring, adds or deletes a directed edge or reverses the direction of an existing directed edge, to produce the directed acyclic graph that gives the maximum score at each stage. The available edge set for the edge orientation phase is restricted to those of the skeleton located by the MMPC algorithm.  A list is kept of the previous 100 structures explored and only those add / delete / reverse operations that do not result in a DAG on the list are considered. When 15 changes are made to the DAG without an increase in the maximum score that has been encountered up to that point in the search, the algorithm terminates and delivers the structure with the highest score obtained so far.  

The standard scoring functions are those obtained by assuming that, given the graph structure, the probability distribution may be factorised according to the graph, with a  Dirichlet distribution over the parameter space. If a uniform prior distribution is taken over the set of possible graph structures, the score function becomes the likelihood function for the data given the graph structure. This is described in Heckerman, Geiger and Chickering~\cite{HGC} and is the method used in the MMHC algorithm. 

Other score functions are the Bayesian Information Criterion (BIC) (Schwartz~\cite{Sc}), the Akaike Information Criterion (AIC) (Akaike~\cite{Ak}), Minimum Description Length (MDL) (Rissanen~\cite{Ris1} and~\cite{Ris2}) and the K2 scoring method (Cooper and Herskovitz~\cite{CH}).  The Bayesian Dirichlet scoring method has the following necessary property: Markov equivalent structures have the same score.

The most basic search-and-score algorithm is the Greedy Search, which searches the space of Directed Acyclic Graphs (DAGs) (Chickering, Geiger and Heckerman~\cite{CGH}). The size of the search space is super-exponential in the number of variables and there have been several modifications to reduce the size of the search space. The Greedy Bayesian Pattern Search algorithm (GBPS) searches over equivalence classes of DAGs. This is basically a search over {\em essential graphs} (definition~\ref{defesgr}). An example is the Greedy Equivalent Search (GES) algorithm (Chickering~\cite{Ch2}).

The K2 algorithm (Cooper and Herskovitz~\cite{CH}) requires an ordering of the variables and uses the K2 metric with a greedy hill - climbing search, where the parent sets for each variable are taken from those variables of a lower index. With this algorithm, it is important to repeat it a large number of times with different orderings. The Sparse Candidate algorithm limits the number of parents that each variable is permitted to have to $k$ and uses a greedy search. The Optimal Reinsertion (OR) algorithm by Moore and Wong~\cite{MW} takes a node, removes all the edges to and from the node, and then re-inserts an optimal set of edges. 

Several constraint based algorithms were compared with the MMPC algorithm in~\cite{TBA}. These decide to delete an edge, thus adding a constraint to the network, based on testing for conditional independence. Among these were the PC algorithm (Spirtes, Glymour and Scheines~\cite{SGS}) and The Three Phase Dependency Analysis (TBDA) by Cheng, Greiner, Kelly, Bell and Liu~\cite{CGKBL}, which uses an information score function when testing for conditional independence. 

The CB algorithm by Singh and Valtorta~\cite{SV} uses a search-and-score algorithm developed by Spirtes Glymour and Scheines, the SGS algorithm (which was the prototype of their PC algorithm) to order the nodes and then uses the K2 algorithm to orient the edges. The PC+GBPS algorithm, by Spirtes and Meek~\cite{SM} uses the PC algorithm to establish an initial pattern, which is then used as the basis of a search and score procedure. The Essential Graph Search (EDS) by Dash and Druzdzel~\cite{DD} uses the PC algorithm repeatedly, with random node orderings and thresholds. The BENEDICT method, which comes from BElief NEtworks DIscovery using Cut - set Techniques is another method that defines a search-and-score metric to measure the distance between a proposed graphical model and the empirical data, using $d$-separation to localise the problem into smaller sets of variables.

Kovisto and Sood~\cite{KS} developed an exact algorithm to identify the network that maximises the posterior probability density, using Bayesian Dirichlet scoring. This algorithm works well for data sets with less than order 100 variables, but cannot be scaled to the large data sets, containing several thousand variables, that are required for learning gene regulatory pathways.

There are several other variants in the literature on the same basic ideas. Goldenberg and Moore~\cite{GM} presented an algorithm for learning Bayesian networks from sparse data  by locating sets that occur frequently; their approach may be used on large variable sets.

There are also variational methods, discussed in Jordan,   Ghahramani,  Jaakkola and Saul~\cite{JGJS}, which typically cannot be used for large networks.

\section{Evaluation} Algorithms are usually compared against a procedure known as the {\em gold standard}.

\begin{Defn}[Gold Standard] The {\em gold standard}  is the procedure where all possible DAGs are enumerated and scored  using an appropriate score function.
\end{Defn} 

\noindent In~\cite{TBA}, several networks were used to provide a comprehensive summary of the performance of MMHC against the following algorithms:
\begin{itemize}
 \item Sparse Candidate (SC by Friedman, Nachman and Pe'er, 1999~\cite{FNP}
\item PC (by Spirtes, Glymour and Scheines, 2000~\cite{SGS})
\item Three Phase Dependency Analysis (TBDA, Cheng et. al., 2002~\cite{CGKBL})
\item Optimal Reinsertion (OR, Moore and Wong, 2003~\cite{MW})
\item Greedy Equivalent Search (GES, Chickering 2002,~\cite{Ch2})
\end{itemize}
In this article, the $M^3PC$ algorithm is compared with the
Maximum Minimum Hill Climbing Algorithm (MMHC)  in ~\cite{TBA}.  The main functions of this article are as follows.
\begin{enumerate} 
\item Determine the immoralities in addition to the skeleton using the MMPC stage of the algorithm, thus rendering the edge orientation phase of the MMHC algorithm (which is the computationally expensive phase) unnecessary. It is clear from the listing of the algorithm that this modification can be incorporated into the MMPC algorithm of~\cite{TBA} with little computational overhead, thus leading to a major reduction in the computational time necessary.
\item Point out the difficulties with the method for determining conditional independence that became immediately apparent in a simple example on six variables,  suggest a modification  and   justify the  modification  theoretically.  

\item In addition, algorithm~\ref{MMPCcheck} is inserted to deal with some cases when the conditional independence statements do not correspond to a probability distribution that has a faithful Bayesian network. The example in subsection~\ref{subsWAM}, where the assumption of faithfulness does not hold, indicates that an algorithm should be able to deal with this.
\end{enumerate}

\noindent The main purpose of the algorithm is to find a suitable graphical model for the dependency structure among large numbers of variables. The main application in view is genetic data for locating gene regulatory pathways.

\subsection{Women and Mathematics}\label{subsWAM} 
\begin{figure}  
\begin{center} 
\[ \UseTips \xymatrix{ 
*++[o][F]{B}\ar[r] \edge{dr} & *++[o][F]{E} &   *++[o][F]{C}\ar[d] \ar[l]\\
   *++[o][F]{A} &  
 *++[o][F]{D}\ar[r] \ar[u]  &
*++[o][F]{F}  } \]
\end{center}
\caption{Women and Mathematics - Essential graph obtained using $M^3$PC algorithm}\label{WAMfigure2} 
\end{figure}
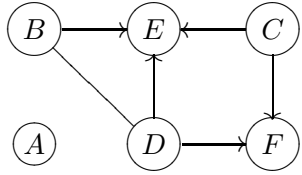

The algorithm was applied to the `Women and Mathematics' data set. This data set is instructive, for several reasons:
\begin{itemize}
\item It motivates the   modification for determining conditional independence in the $M^3PC$ algorithm, the procedure for determining whether a statement $X \perp Y|S$ should or should not be added to the set of conditional independence statements used to construct the graph. Without the modification, there is a missing edge, the omission of which gives a misleading impression of the dependence structure between the variables.
\item It motivates the last stage (algorithm~\ref{MMPCcheck}) of the algorithm, since there is not a distribution with a faithful graphical representation corresponding to the logically consistent set of dependence relations derived using this method for determining the dependence relations. 
\item Even with the modification, there are $d$-separation statements in the final graph that do not correspond to independence statements. In figure~\ref{WAMfigure2}, $E \perp F \|_{\cal G} \{C,D\}$, but $G(E,F|C,D) = 9.83$ (where $G$ is defined by equation (\ref{eqtestgee})) while $\chi_{4,0.05} = 9.45$, so that conditional independence is {\em rejected} at the $5\%$ significance level. This shows the limitations of the algorithm, even with the modification, and the problem arises with the assumption of faithfulness in determining edge selection. Broadly speaking, a situation where $A$ and $C$ are independent, $B$ and $C$ are independent, but where $A$ and $B$ taken together tell us everything about $C$, will not be detected by an algorithm that selects edges based on an assumption that there is a faithful distribution.
\end{itemize}

The variables are $A$- lecture attendance (whether the person attended a special lecture by a female mathematician designed to encourage women to take up mathematics) $B$ - gender $C$- school type (urban / suburban), $D$ - `I'll need mathematics in my future work' (agree = y / disagree = n), $E$- subject preference (mathematical sciences / liberal arts), $F$ - future plans (higher education / immediate job). The data is given in table~\ref{WAMtable}.

\begin{figure}  
\begin{center} 
\[ \UseTips \xymatrix{ 
*++[o][F]{B}\edge{r} \edge{dr} & *++[o][F]{E}\edge{d}&   *++[o][F]{C}\ar[d]\\
   *++[o][F]{A} &  
 *++[o][F]{D}\ar[r]   &
*++[o][F]{F}  } \]
\end{center}
\caption{Women and Mathematics - Essential graph obtained using MMPC algorithm, plus testing of vee - structures to locate immoralities}\label{WAMfigure} 
\end{figure}
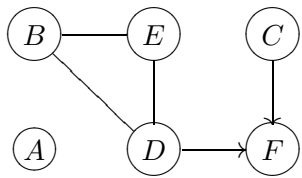

The MMPC from~\cite{TBA} (that is, without any modifications) produced the skeleton corresponding to the graph in figure~\ref{WAMfigure}. This in line with other algorithms; the   $(MC)^3$ and $A(MC)^3$ Markov chain Monte Carlo algorithms by~\cite{MAPV} produce the essential graph of figure~\ref{WAMfigure}. 

But this graph does not represent the dependence relations in the data; $H_0 : C \perp E$ is {\em rejected} when a significance level of $5\%$ is employed, which is the nominal significance level used by the MMPC algorithm, while $C$ is $d$-separated from $E$ in figure~\ref{WAMfigure}.

The $p$ value is slightly greater than $1\%$; the hypothesis test would not be rejected at the $1\%$ significance level. The fact that there is no edge between $C$ and $E$ in the graph returned by the   $A(MC)^3$ algorithm is due to the penalty for additional edges in the score function. 

The MMPC algorithm proceeds as follows: with the ordering $A,B,C,D,E,F$ on the variables, the stages are as follows:

\paragraph{Stage 1} The candidate neighbour sets after stage 1 are
\[{\cal Z}_A = \phi, \quad {\cal Z}_B = \{D,E\}, \quad {\cal Z}_C = \{E,F\}, \quad {\cal Z}_D = \{B,E,F\}, \quad {\cal Z}_E = \{B,C,D\}, \quad {\cal Z}_F = \{B,C,D\}.\]

\paragraph{Stage 2} This involves the removal of nodes $X_j$ from ${\cal Z}_k$ if $X_k \not \in {\cal Z}_j$. This only results in one change; $B$ is removed from ${\cal Z}_F$, giving 

\[{\cal Z}_A = \phi, \quad {\cal Z}_B = \{D,E\}, \quad {\cal Z}_C = \{E,F\}, \quad {\cal Z}_D = \{B,E,F\}, \quad {\cal Z}_E = \{B,C,D\}, \quad {\cal Z}_F = \{C,D\}.\]

\paragraph{Stage 3} In this stage, node $X_k$ is removed from the set ${\cal Z}_j$ if there is a subset $S$ of the current set ${\cal Z}_j \backslash \{X_k\}$ such that $X_j \perp X_k | S$. There is only one alteration at this stage; node $C$ is removed from ${\cal Z}_E$. It turns out that $C \not \perp E$ and $C \not \perp E | D$, but that $C \perp E | \{B,D\}$. After this stage, the sets of neighbours are

\[{\cal Z}_A = \phi, \quad {\cal Z}_B = \{D,E\}, \quad {\cal Z}_C = \{E,F\}, \quad {\cal Z}_D = \{B,E,F\}, \quad {\cal Z}_E = \{B, D\}, \quad {\cal Z}_F = \{C,D\}.\]

\paragraph{Stage 4} This stage makes a final removal of nodes $X_k$ from ${\cal Z}_j$ if $X_j \not \in {\cal Z}_k$. Only one alteration is made; $E$ is removed from ${\cal Z}_C$. The final neighbour sets are 

\[{\cal Z}_A = \phi, \quad {\cal Z}_B = \{D,E\}, \quad {\cal Z}_C = \{F\}, \quad {\cal Z}_D = \{B,E,F\}, \quad {\cal Z}_E = \{B, D\}, \quad {\cal Z}_F = \{C,D\}.\]

\noindent Finally, testing vee - structures (definition~\ref{defveestr}) to see if they are immoralities (definition~\ref{defimm}) gives: $E \not \perp F$ and $E \perp F | D$ so that $(E,D,F)$ is not an immorality. $D \perp C$, but $D \not \perp C | F$ so that $(D,F,C)$ is an immorality. The essential graph corresponding to the dependence / independence structure returned by the MMPC algorithm, with the vee structures tested to determine whether or not they are immoralities, is therefore given in figure~\ref{WAMfigure}.

\begin{table}\label{WAMtable}
\begin{tabular}{lcc|rrrrrrrr} \hline &&school&\multicolumn{4}{c}{suburban}&\multicolumn{4}{c}{urban}\\
&&  gender&\multicolumn{2}{c}{female}&\multicolumn{2}{c}{male}&\multicolumn{2}{c}{female}&\multicolumn{2}{c}{male}\\&&lecture&y&n&y&n&y&n&y&n\\ \hline
future & preference & `need mathematics' & &&&&&&&\\
college & mathematical & y &37 & 27 & 51 & 48 & 51 & 55 & 109 & 86 \\
&  &  n & 16 & 11 & 10 & 19 & 24 & 28 & 21 & 25 \\
& arts & y & 16 & 15 & 7 & 6 & 32 & 34 & 30 & 31 \\
&& n & 12 & 24 & 13 & 7 & 55 & 39 & 26 & 19 \\
job & mathematical & y & 10 & 8 & 12 & 15 & 2 & 1 & 9 & 5 \\
&&n& 9 & 4 & 8 & 9 & 8 & 9 & 4 & 5 \\
& arts & y & 7 & 10 & 7 & 3 & 5 & 2 & 1 & 3 \\
&&n & 8 &4 & 6 &4 & 10 & 9 & 3 & 6 \\ \hline
\end{tabular}

\caption{Data for `Women in Mathematics' Source: Lacampagne (1979~\cite{Lac}} 
 
\end{table}

This graph does not represent the conditional dependencies that have been established. At nominal significance level of $5\%$, the significance level being used, the test rejected $C \perp E$, but in the graph in figure~\ref{WAMfigure}, $C$ is $d$-separated from $E$. For the pair of variables $C,E$, the cell counts for $n_{CE}$ are 
\[n_{CE}(1,1) = 194, \quad n_{CE}(1,0) = 149, \quad n_{CE}(0,1) = 439, \quad n_{CE}(00) = 305\]
so that
\begin{eqnarray*}\lefteqn{G^2_{CE} = 2 \times \left(294\log\frac{294\times 1187}{443 \times 733} + 149 \log \frac{149\times 1187}{443\times 454}\right.}\\&& \hspace{20mm} \left.  + 439 \times \log\frac{439\times 1187}{774 \times 733} + 305 \times \log\frac{305\times 1187}{454\times 744}\right) = 6.42.\end{eqnarray*}
\noindent Note that
\[ \chi^2_{1,0.05} = 3.84, \qquad \chi^2_{1,0.01} = 6.63\]
\noindent so the hypothesis that $C \perp E$ is rejected at the $5\%$ significance level; the test has a $p$ value close to $0.01$. Yet, for the network in figure~\ref{WAMfigure}, $C \perp E \|_{\cal G} \phi$ ($C$ is $d$-separated from $E$ when none of the variables are instantiated).

Similarly, in the graph, $C \perp E \|_{\cal G} \{D\}$. But the hypothesis that $C \perp E | D$ is {\em rejected} at the $5\%$ significance level, so that $C \not \perp E | D$ is {\em accepted}. 

\[n_{CDE}(1,1,1) = 208, \quad n_{CDE}(1,1,0) = 71, \quad n_{CDE}(1,0,1) = 86, \quad n_{CDE}(1,0,0) = 78\]
\[n_{CDE}(0,1,1) = 318, \quad n_{CDE}(0,1,0) = 138, \quad n_{CDE}(0,0,1) = 124, \quad n_{CDE}(0,0,0) = 167.\]
\[n_D(1) = 735,\quad n_D(0) = 455\]
\[n_{CD}(1,1) = 279, \quad n_{CD}(1,0) = 164, \quad n_{CD}(0,1) = 456, \quad n_{CD}(0,0) = 291\]
\[n_{DE}(1,1) = 526, \quad n_{DE}(1,0) = 209, \quad n_{DE}(0,1) = 210, \quad n_{DE}(0,0) = 245\]

\noindent so that
\[G(C,E|D)   = 2 \times \sum_{d=0}^1 \sum_{c,e}   n_{CDE}(c,d,e)\log\frac{n_{CDE}(c,d,e)n_D(d)}{n_{CD}(c,d)n_{DE}(d,e)} = 6.063 \geq 5.99 = \chi^2_{2,0.05}.\]

\noindent  The hypothesis that $C \perp E | D$ is rejected, although not by much; the $p$ value is only slightly less than  $5\%$.

The problem here is not the lack of a faithful graphical model. Rather, it is that the conditional independence statements that have been accepted are inconsistent with the conditional independence statements that have been rejected. By theorem~\ref{thci1}, 
\[C \perp E | \{B,D\} \quad \mbox{and} \quad \{B, D\} \perp C   \Rightarrow C \perp E\]
\[C \perp E | \{B,D\} \quad \mbox{and} \quad  B \perp C | D \Rightarrow C \perp E | D.\]

\noindent Now, the individual hypothesis tests $H_0: B \perp C$, $H_0: D \perp C$ and $H_0: B \perp C | D$ each result in $H_0$ not rejected. But $C \not \perp E$ is accepted and $C \not \perp E | D$ is also accepted. Therefore, the statement $C \not \perp E | \{B,D\}$ should also be accepted, otherwise inconsistencies appear in the independence structure, leading to a graph that indicates $d$-separation where conditional dependence has been established. Any undergraduate engineering student will fail their introductory course in statistics if they have not learned that `do not reject the null hypothesis' does not mean `accept the null hypothesis'; this example provides an illustration of the importance of this simple basic principle. \vspace{5mm}

\noindent When the modification in the way that conditional independence is established is used, the final sets of neighbours are 

\[{\cal Z}_A = \phi, \quad {\cal Z}_B = \{D,E\}, \quad {\cal Z}_C = \{E,F\}, \quad {\cal Z}_D = \{B,E,F\}, \quad {\cal Z}_E = \{B, C, D\}, \quad {\cal Z}_F = \{C,D\}.\]

\noindent When the  modification to locate the immoralities is introduced,  there are five vee structures to be checked: $B-E-C$, $C-E-D$, $E-D-F$, $C-F-D$, $E-C-F$ . Since $B \perp C$, the algorithm declares $(B,E,C)$ is an immorality. Since  $C  \perp D$,  the proposed algorithm declares that $C-E-D$ is an immorality. Since $E \perp F | \{B,C,D\}$ (from stage 1 of the algorithm), it follows that $E-D-F$ is {\em not} an immorality, since $D$ is in the conditioning set. Since $C \perp D$, it follows that $C-F-D$ is an immorality. Since $E \perp F | \{B,C,D\}$, it follows that $E-C-F$ is not an immorality. Once the immoralities have been located, the only remaining undirected edge is $B-D$. Since it does not belong to any of the structures in figure~\ref{strongprotect}, it remains undirected. The essential graph obtained is given in figure~\ref{WAMfigure2}.  

\paragraph{A Note on Faithfulness} It is important to note that the set of conditional independence relations constructed by the $M^3PC$ algorithm, with the modification to ensure that the conditional independence statements accepted are consistent with those rejected, do not correspond to a distribution that has a faithful graphical model. In the essential graph in figure~\ref{WAMfigure2}, $B \not \perp C \|_{\cal G} E$. That is, $B$ is not $d$-separated from $C$ when $E$ is instantiated. But 
\[G(B,C;E) = 0.617 < 5.99 = \chi^2_{2;0.05}\]
so that the data does {\em not} provide evidence for  $B \not \perp C | E$.

There is a very significant conditional dependence between $C$ and $D$ given $E$;
\[G(C,D;E) = 56.57 > 5.99 = \chi^2_{2;0.05}\]
which is reflected in the immorality $C-E-D$ in the graph, while $C \perp D$. 

\paragraph{Another note on faithfulness} As mentioned earlier,

\[G(E,F|C,D) = 9.83 > 9.45 = \chi^2_{4;0.95}\]
\noindent so that $E \not \perp F | \{C,D\}$ is {\em accepted} at the $5\%$ significance level. Nevertheless, in figure~\ref{WAMfigure2}, $E \perp F \|_{\cal G} \{C,D\}$; any trail between $E$ and $F$ has to pass through an instantiated fork connection. 

\subsection{Comparisons with the MMHC Algorithm} An extensive comparison of the MMHC algorithm against other structure learning algorithms is made in~\cite{TBA} and the reader is referred to~\cite{TBA} for these comparisons. With reference to these results, it is of interest to understand where the modifications proposed for the  $M^3PC$ algorithm would influence the results. 

As noted earlier, the MMHC algorithm is carried out in two parts; firstly the MMPC algorithm to locate the skeleton and then the edge orientation phase to produce a directed acyclic graph. The edge orientation phase uses the edges in the skeleton and at each step the algorithm modifies a single edge. The algorithm starts with the empty graph. From the entire set of  possible edges, which are those located by the MMPC algorithm, it chooses the add edge / delete edge / reverse direction operation that produces a network that has not been considered previously, with the greatest increase in the BDeu score function. This continues until there have been $15$ steps without any increase in the maximum score function encountered. The algorithm returns the network corresponding to the maximum score function encountered. 
Each operation will change the parent set of at most two variables. Add edge and delete edge changes the parent set of a single variable, reverse edge changes the parent sets of two variables. The updated score function when a single edge is obtained by updating the previous score function and requires at most one statistical call to the data set.

Suppose the skeleton contains $S$ edges. If each edge, with an orientation that corresponds to a directed acyclic graph with the correct essential graph were located at each step, so that no delete edge or reverse edge operations were necessary, the total number of statistical calls necessary would be bounded above by $2S^2$, assuming that the algorithm stops as soon as all the edges are added. The number of steps required by the MMHC algorithm is usually substantially larger   than $2S^2$. 

For each edge not present, it may be added in one of two ways; for each edge already present, it may either be deleted or re-oriented. The algorithm works by checking all the `legal' possibilities, i.e. those that result in a directed graph without cycles.

An easy {\em lower} bound on the number of calls, assuming that one call is required for each iteration, is given by $S^2$.

The following `back of an envelope' calculation gives a ball park figure for the number of calls in terms of the size of the parent sets. Let $\tilde{\Pi}$ denote the average parent set size, then the number of edges is given by $S =  \tilde{\Pi} d$; if $\tilde{\Theta}$ is the average size of the neighbour sets for each neighbour, then there is a lower bound of $\frac{1}{4}\tilde{\Theta}^2 d^2$ on the number of statistical calls. 

It is pointed out in~\cite{TBA} that for the test networks used in that article, it is usual that $\tilde{\Theta} \sim O(\tilde{C})$, where $\tilde{C}$ is the average size of the neighbour sets after the first stage of the $M^3PC$ algorithm. The {\em lower} bound for the edge orientation phase is of order $\frac{1}{4}\tilde{C}^2 d^2$, while an upper bound for the number of calls for the MMPC algorithm, with a modification to locate the immoralities, is $d(d-1)  + e C^l d$ (given by equation (\ref{expmmpc})) and an upper bound for the $M^3PC$ (both modifying the method to determine independence and also locating the immoralities) is given by $d(d-1)\frac{C^l}{l!}(l+2)2^{l-1}$ (equation (\ref{equppbd})), where $C$ is the size of the largest neighbour set and $l$ is the size of the largest conditioning set used.

If the complexity is measured by the number of statistical calls, where a statistical call means a operation that involves using the entire data matrix, then the upper bound for the complexity of the $M^3PC$ algorithm is $d(d-1)$ calls, and this is the same value,  whether or not the modifications are employed.

\paragraph{Evaluation results from~\cite{TBA}}  The article~\cite{TBA} evaluates the MMPC algorithm using test networks   found on 
 
\noindent {\tt  http://www.cs.huji.ac.il/labs/compbio/Repository/networks.html}

\noindent These are often used in evaluation, to provide an objective standard for assessing the performance of a structure learning algorithm.  In~\cite{TBA}, section 9.2.7, several examples are given of the performance of the Minimum Maximum Hill Climbing algorithm when applied to the `ALARM' network, one of the networks on the web page above, and various networks produced by tiling the ALARM network. That is, to get a larger network, the ALARM network is repeated many times, with suitable edges added to connect the various tiles. Some results are given for:
\begin{itemize}
 \item a single copy of ALARM (37 variables and 46 edges) 
\item ALARM10 - 10 tiles of ALARM
\item 135 tiles of ALARM (4995 variables and 6845 edges)
\item 270 tiles of ALARM (9990 variables and 13640 edges)
\end{itemize}
 
The results in~\cite{TBA} for the third of these, with 4995 variables are instructive, because it is the only place in that article where separate  indications are given of the time taken for the MMPC stage of the algorithm and the edge orientation stage of the algorithm. 

Only the MMPC algorithm is applied to the network with 9990 variables.

The number of missing edges and extra edges are given for the single copy and the tiling with 10 copies.

For a single copy of the ALARM network, the sizes of the parent sets and neighbour sets, and the frequency with which these size occurs, are given in table~\ref{tblALARMfreq}.

\begin{table}
 \[\begin{array}{|cc|cc|} \hline \mbox{size of parent sets} & \mbox{frequency } & \mbox{size of neighbour sets} & \mbox{frequency}\\ \hline 
    0 & 12 &1 & 10\\
1 & 8 & 2 & 12\\ 
2 & 14 & 3 & 8\\
3 & 2 &4 & 3\\
4 & 1 & 5 & 3\\
&&6 & 1 \\ \hline 
   \end{array}\]
\caption{Sizes of parent sets for the ALARM network}
\label{tblALARMfreq}
\end{table} 

For the example with $4990$ variables, the MMPC stage took 19 hours, while the edge orientation stage took approximately 300 hours. That is, the edge orientation stage of the algorithm took 15 times as long as the skeleton reconstruction phase. The complexity of the $M^3PC$ algorithm is similar to the complexity of the MMPC algorithm of~\cite{TBA}, so the computational advantage is clear.

For the network with $9990$ variables and $13640$ edges, only $1000$ training examples were used. The time taken was reported in~\cite{TBA} as $62$ hours. Compared with the amount of time taken by the network on $4995$ variables, the ratio of the time taken for the ratio of small: large network  is $1:3.26$, which is considerably less than $1:4$, the ratio expected if the time taken is proportional to $d(d-1)$. This  may well be because the four parallel processors were working faster than four times the speed of a single processor. The network with $5000$ variables was reconstructed using a single processor, while the network with $10000$ variables was reconstructed using four parallel processors. There may have been proportionately fewer calls, because fewer training examples were used, so that fewer test statistics could be computed with accuracy. 
 
\paragraph{Accuracy results from~\cite{TBA}} 
For the various networks, the accuracy of reconstruction was as follows:
\begin{enumerate}
\item The ALARM10 network (370 variables, 500 edges) Based on repeated trials, each with $5000$ training examples, the reconstruction in~\cite{TBA} had an average of $27.2$ extra edges and $109.0$ missing edges.  

\item The network constructed from 135 tiles of the ALARM network (4995 variables and 6845 edges) Based on a single run, using $5000$ training examples, the reconstruction had 1340 extra edges and 1076 missing edges.  

\item The network constructed from 270 tiles of the ALARM network (9990 variables and 13640 edges) Based on a single run, using $1000$ training examples, the reconstruction had $11068$ extra edges and  $2572$ missing edges.  
\end{enumerate}

\noindent The networks were used to generate random data according to the distribution. The algorithm was then applied to the randomly generated data to reconstruct the network.

\paragraph{Extra Edges} Although the reconstructed networks for $370$, $4995$ and $9990$ variables all contain edges additional to the underlying graph structure of the probability distribution used to generate the data, this reflects the data rather than the algorithm. The algorithm rejects an edge if the direct association is insignificant at the $5\%$ level, or if any of the conditional associations computed are insignificant at the $5\%$ level. Therefore, edges appearing represent associations genuinely present in the data, even if the associations were not present in the underlying distribution from which the data was generated.

Consider $d$ independent variables. Suppose that, in the data generated, two variables are associated with probability $p$. Let $X$ denote the number of associations in the data, then $X \sim Bi(\frac{1}{2}d(d-1) , p)$ which, for $\frac{d(d-1)}{2}p \geq 10$ and $p$ small may be approximated as $N(\frac{p}{2}d(d-1), \frac{p}{2}d(d-1))$. If $d = 10000$ and $p = 2\times 10^{-4}$, then
\[p(X \geq 10000) \simeq 0.5 \qquad\mbox{and}\qquad p(X \geq 9700) \simeq 0.99.\]

\noindent The numbers of `extra edges' quoted by~\cite{TBA} in their results would be consistent with the number of genuine associations that may be expected in data if the probability of an association appearing in the data between two independent variables is of order $10^{-4}$. The assumption of faithfulness ensures the rejection of edges unless a clear association is shown; the method of accepting conditional independence statements increase the probability of rejecting an edge even when there is a clear association in the data. 

\paragraph{Missing Edges} The results for the ALARM networks with $500$, $6845$ and $13640$ edges   in~\cite{TBA} show $109$ (averaged over several trials), $1340$ and $2572$ missing edges respectively.  This is a substantial number, representing $22\%$, $20\%$ and $19\%$ of the edges respectively in the underlying distributions from which the data is obtained. This suggests that the edges are missing for the same reason that the edge between $C$ and $E$ is rejected in the example in subsection~\ref{subsWAM}; conditional independence tests with larger numbers of variables fail  to reject a hypothesis and thus imply independence for a smaller conditioning set where independence has already been rejected.

\section{Conclusion} The Maximum Minimum Hill Climbing  algorithm, introduced by Tsamardinos, Brown and Aliferis is a powerful algorithm for locating the skeleton of a faithful graph, if there exists a faithful Bayesian network, for large   networks, where the dependency structure is sparse. This article takes the first stage of the MMHC algorithm, the stage where the skeleton is determined, and locates the immoralities without requiring additional statistical calls. This enables the essential graph to be constructed, which captures all the Markov properties that are contained in any faithful directed acyclic graph.    Its implementation is feasible whenever the implementation of the MMPC algorithm of Tsamardinos, Brown and Alifeis is feasible and it leads to a substantial decrease in the complexity compared with the MMHC algorithm.

The method for determining conditional independence statements used in the MMPC algorithm may lead to inconsistencies; an independence statement that is not rejected is accepted, and may contradict earlier dependence statements that have been established. This difficulty appears in the WAM data, discussed in subsection~\ref{subsWAM}. A method for producing a set of independence statements that are consistent with each other is therefore proposed. 

The conditional dependence / independence relations established may not have a faithful graphical representation. This is motivated by the WAM data set of subsection~\ref{subsWAM}, a randomly chosen data set of $1000$ observations on 6 binary variables, which exhibits conditional dependence / independence relations that do not have a faithful graphical representation. An economical method for locating an essential graph with minor modification to the skeleton is suggested that returns an essential graph containing all the dependencies derived for the construction of the skeleton, although this is not exhaustive; if the faithfulness assumption does not hold, the procedure may miss some dependencies that were not established in the skeleton construction phase.

The value of the first of these modifications is clearly gives a substantial reduction of the time required. The second leads to greater accuracy, dealing with a situation motivated by the WAM example (subsection~\ref{subsWAM}). The third of these, inserting algorithm~\ref{MMPCcheck} is necessary if the faithfulness assumption is not guaranteed.  

The WAM was a randomly chosen data set, chosen simply for its convenience as a six variable set that had already been used to test other algorithms and not because it was imagined that it might illustrate lack of faithfulness. This indicates that lack of faithfulness is not rare and that  situations requiring the modification in the procedure for determining conditional independence are not rare.

Unfortunately, while programming the algorithm proposed in this article is straightforward enough, and while initial results look  promising, the author did not have sufficient computational power available to engage in larger experiments. The additional time taken when the modification for determining conditional independence is included is therefore unknown for larger networks.  The purpose of the article was to a) point out  the theoretical point that the computations made in the MMPC algorithm, if used properly, already give the immoralities and hence Markov structure of the network, thus rendering the edge orientation phase of the MMHC algorithm unnecessary, b) point out the theoretical problem that could occur with the method for determining conditional independence, show that it does happen in a straightforward real data set and propose a solution that represents all associations established from data and c) point out, using a real data set, that lack of faithfulness occurs in practise and therefore has to be considered in a structure learning algorithm.  

\section{Appendix}\label{app} The appendix describes the background material so that the article is self contained. Let $V = \{X_1, \ldots, X_d\}$ denote a collection of random variables, each with a multinomial distribution, which are related, with joint probability distribution $p$. In the article, ${\cal G} = (V,E)$ is used to denote a graph with variable set $V$ and edge set $E$. In general, $E = D \cup U$ where $D$ is the set of directed edges and $U$ is the set of undirected edges. If the graph is directed, the notation ${\cal G} = (V,D)$ will be used and if undirected, the notation ${\cal G} = (V,U)$ will be used. The notation  $(X_i, X_j)$ is used to denote a directed edge between two variables $X_i$ and $X_j$. The notation $\langle X_i, X_j \rangle$ is used to denote an undirected edge between two variables. In diagrams, the notation
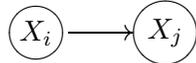
\[  \UseTips \xymatrix{ 
*++[o][F]{X_i} \ar[r] & *++[o][F]{X_j}  }\]
denotes $(X_i, X_j)$, a directed edge from $X_i$ to $X_j$ and  
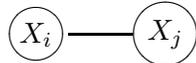
\[  \UseTips \xymatrix{ 
*++[o][F]{X_i} \edge{r} & *++[o][F]{X_j}  }\]
denotes $\langle X_i , X_j \rangle$, an undirected edge between $X_i$ and $X_j$. 

For sets of random variables $A$, $B$ and $C$, $A \perp B$ denotes that $A$ is independent of $B$ and $A \perp B | C$ denotes that $A$ is conditionally independent of $B$ given $C$. $A \not \perp B$ denotes that $A$ is not independent of $B$ and $A \not \perp B | C$ denotes that $A$ is not conditionally independent of $B$ given $C$. 

\begin{Defn}[Directed Acyclic Graph] A graph ${\cal G} = (V,D)$ where $V$ is the set of nodes and $D$ the set of edges is said to be a directed acyclic graph if each edge in $D$ is directed and for any node $X \in V$ there does not exist a distinct set of nodes $Y_1,\ldots,  Y_n$ for $n \geq 1$ such that $X \neq Y_i$ for all $i = 1, \ldots, n$, the edge $(X , Y_1) \in D$, $(Y_n, X) \in D$ and $(Y_i, Y_{i+1}) \in D$ for $i = 1, \ldots, n-1$. 
 \end{Defn}

\noindent There are several possible terminologies for the various connections in the following definition; this is one of the standard ones.

\begin{Defn}[Fork connection, Chain connection, Collider connection]\label{deffcc}
   Two variables $X_1$ and $X_3$ are connected by a fork connection, via a third variable $X_2$ if the structure is
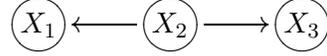
\[ \UseTips \xymatrix{ 
*++[o][F]{X_1}  & *++[o][F]{X_2}\ar[l] \ar[r] &   *++[o][F]{X_3}  }\]
$X_1$ and $X_3$ are connected by a chain connection via $X_2$ if the structure is
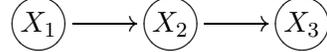
\[ \UseTips \xymatrix{ 
*++[o][F]{X_1} \ar[r] & *++[o][F]{X_2} \ar[r] &   *++[o][F]{X_3}  }\]
$X_1$ and $X_3$ are connected by a collider connection via $X_2$ if the structure is
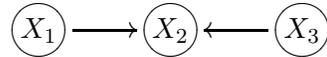
\[ \UseTips \xymatrix{ 
*++[o][F]{X_1}\ar[r]  & *++[o][F]{X_2}  &   *++[o][F]{X_3}\ar[l]  }\]
The nodes at the centre of a fork, chain or collider connection may be referred to as fork, chain or collider nodes. 
\end{Defn}

\begin{Defn}[Trail]\label{deftrail} A {\em trail} between two nodes $X$ and $Y$ is a collection of nodes  $Z_1, \ldots, Z_n$ such that, setting $X= Z_0$ and $Y = Z_{n+1}$, for $i = 0,\ldots, n$, $(Z_i, Z_{i+1})$ or $(Z_{i+1}, Z_i)$ or $\langle Z_i, Z_{i+1} \rangle$ are in $E$. 
 \end{Defn}

\begin{Defn}[Directed Path] \label{defdrpath} A {\em directed path} between two nodes $X$ and $Y$ is a collection of nodes  $Z_1, \ldots, Z_n$ such that, setting $X= Z_0$ and $Y = Z_{n+1}$, for $i = 0,\ldots, n$, $(Z_i, Z_{i+1}) \in D$. 
 \end{Defn}

\begin{Defn}[Descendant, Ancestor, Parent, Child]\label{defpcad}
In a directed acyclic graph, $X$ is an ancestor of $Y$ if there is a directed path from $X$ to $Y$. The node $Y$ is a descendant of $X$ if there is a directed path from $X$ to $Y$. The node $X$ is a parent of $Y$ if there is a directed edge $(X, Y) \in D$. The node $Y$ is a child of $X$ if there is a directed edge $(X, Y) \in D$. 
\end{Defn}

\begin{Defn}[$S$-active trail, blocked trail]\label{defbltr} 
 Let ${\cal G} = (V, D)$ be a directed acyclic graph. Let $S \subset V$. A trail between two variables $X$ and $Y$ not in $S$ is said to be $S$-active  if
\begin{enumerate}
 \item Every collider node $Z$ is in $S$ or has a descendant in $S$.
\item Every other node is outside $S$.
\end{enumerate}
A trail between $X$ and $Y$ that is not $S$-active is said to be blocked by $S$. 
\end{Defn}
\begin{Defn}[Instantiated] \label{definst} Let $V = \{X_1, \ldots, X_d\}$ denote a set of random variables. A random variable $X_j$ is said to be {\em instantiated} when the state of the variable is known.
\end{Defn}

\begin{Defn}[$d$-separation] \label{defdsep} 
Let ${\cal G} = (V,D)$ be a directed acyclic graph. Let $A$, $B$ and $S$ be three disjoint subsets of $V$. Two nodes $X_i$ and $X_j$ are said to be $d$-separated given $S$ if all trails between $X_i$ and $X_j$ are blocked by $S$. This is written $X_i \perp X_j \|_{\cal G} S$. Two sets $A$ and $B$ are said to be $d$-separated by $S$ if for any $X_i \in A$ and $X_j \in B$, $X_i \perp X_j \|_{\cal G} S$.  This is written  $A \perp B\|_{{\cal G}} S$. 
\end{Defn}

\begin{Defn}[Bayesian Network, Factorisation] \label{defBN} A Bayesian network is a pair $({\cal G}, p)$, where ${\cal G} = (V, D)$ is a directed acyclic graph, $V$ is the node set and $D$ is the set of directed edges such that
 \begin{itemize}
\item For each node $X_v \in V$ with no parent variables, there is assigned a probability distribution $p_{X_v}$ and to each node $X_v$ with non - empty parent set $\Pi_v = (X_{b_1^{(v)}}, \ldots, X_{b_m^{(v)}})$ there is  a conditional probability distribution $p_{X_v | \Pi_v}$ such that (declaring $p_{X_v|\Pi_v} = p_{X_v}$ if $\Pi_v = \phi$ the empty set)
\[ p_{X_1, \ldots, X_v} = \prod_{v=1}^d p_{X_v|\Pi_v}.\]
Such a decomposition is known as a {\em factorisation}. 
\item The factorisation is minimal in the sense that $\Pi_j$ is the smallest subset of $\{X_1, \ldots, X_{j-1}\}$ such that $X_j \perp \{X_1, \ldots, X_{j-1}\}\backslash \Pi_j |\Pi_j$.
 \end{itemize}
\end{Defn}

\noindent The definition of $d$-separation is motivated by the following consideration. Consider three variables $X_1$, $X_2$, $X_3$ with joint probability function $p_{X_1,X_2,X_3}$. Suppose that this distribution may be factorised $p_{X_2}p_{X_1|X_2}p_{X_3|X_1}$, according to the fork node. Then $X_1 \not \perp X_3$, but $X_1 \perp X_3 | X_2$. 

If $p_{X_1,X_2,X_3} = p_{X_1}p_{X_2|X_1}p_{X_3|X_2}$, then the factorisation is according to the chain connection. Again, $X_1 \not \perp X_3$, but $X_1 \perp X_3 | X_2$.

If $p_{X_1,X_2,X_3} = p_{X_1}p_{X_3}p_{X_2|X_1,X_3}$, then the Bayesian network is the collider connection. Here $X_1 \perp X_3$, but $X_1 \not \perp X_3 | X_2$. 

These conditional independence statements may be considered as $d$-separation statements in the corresponding directed acyclic graphs, with $S = \{X_2\}$.\vspace{5mm}

If $d$-separation holds, then conditional independence also holds.

\begin{Th}
Let ${\cal G} = (V,D)$ be a directed acyclic graph where $V = \{X_1, \ldots, X_d\}$ is a set of random variables. Let $p$ be a probability distribution that factorizes along ${\cal G}$. Then, for any disjoint subsets $A,B$ and $S$ of $V$, it holds that $A \perp B | S$ if $A \perp B \|_{\cal G} S$. That is, if the $d$-separation statement holds, then $A$ is conditionally independent of $B$ given $S$. 
\end{Th}

\noindent {\bf Proof} This is theorem 2.2 found on page 66 of~\cite{KN}, where a direct proof of the result is given. \qed \vspace{5mm}

The converse does not necessarily hold. When it does, a directed acyclic graph ${\cal G} = (V,D)$ is said to be {\em faithful} to the distribution.

\begin{Defn}[Faithful]\label{deffaith}
 A directed acyclic graph ${\cal G} = (V,D)$ over a variable set $V = \{X_1, \ldots, X_d\}$ is said to be faithful to a probability distribution $p$ over $V$ if for any three disjoint subsets $A, B$ and $S$ of $V$, it holds that 
\[ A \perp B \|_{{\cal G}} S \Leftrightarrow A \perp B | S.\]
\end{Defn}

\begin{Defn}[Immorality]\label{defimm} Let ${\cal G} = (V, E)$ denote a graph, where $E = D \cup U$, $D$ is the set of directed edges and $U$ the set of undirected edges. An {\em immorality} is a triple $(X, Y, Z)$ such that $(X, Y) \in D$, $(Z, Y) \in D$, but $(X , Z) \not \in D$, $(Z , X) \not \in D$ and $\langle X , Z \rangle \not \in U$. 
 \end{Defn}

\begin{Defn}[Skeleton]\label{defskel} 
The {\em skeleton} of a graph ${\cal G} = (V, E)$ where $E = D \cup U$ is the graph obtained by making the graph undirected. That is, the skeleton of ${\cal G}$ is the graph $\tilde{\cal G} = (V, \tilde{E})$ where $\langle X , Y \rangle \in \tilde{E}$ if and only if either $\langle X , Y \rangle \in U$ or $(X , Y) \in D$ or $(Y , X) \in D$.
\end{Defn}

\begin{Defn}[vee - structure]\label{defveestr}
 A {\em vee structure} is defined as a triple of variables $\{X,Y,Z\}$ such that the skeleton contains edges $\langle X,Y \rangle$ and $\langle Y,Z \rangle$, but no edge $\langle X,Z \rangle$.
\end{Defn}

\begin{Defn}[Markov Blanket]\label{defMB}
 In a directed acyclic graph, the Markov blanket ${\cal M}(X)$ of a node $X$ is the set containing the parents of $X$, the children of $X$ and the  parents of children of $X$.
\end{Defn}

\begin{Lmm}\label{lemmMB}
 For any variable $Y \in V \backslash \{\alpha\} \cup {\cal M}(X)$, $X$ is $d$-separated from $Y$ by ${\cal M}(X)$.
\end{Lmm}

\noindent {\bf Proof of lemma~\ref{lemmMB}} Exercise 6 on page 77 of~\cite{KN} \qed \vspace{5mm}

\noindent The following theorem is of crucial importance and comes from Spirtes, Glymour and Scheines~\cite{SGS93}

\begin{Th}\label{thfaithedge}
The skeleton of a faithful graph contains an edge $\langle X, Y \rangle$ between two variables $X$ and $Y$ if and only if $X \not \perp Y | S$ for any set $S \subseteq V \backslash \{X , Y \}$ (including the empty set). 
\end{Th}\vspace{5mm}

\noindent {\bf Proof of theorem~\ref{thfaithedge}} If there exists a set $S$ such that $X  \perp Y | S$ for some set $S$, then all trails between $X$ and $Y$ are blocked by $S$ and hence there is no edge $\langle X , Y \rangle$ in the skeleton of the faithful graph. 

If there is no set $S$ such that $X \perp Y | S$, then from lemma~\ref{lemmMB}, it follows that $Y \in {\cal M}(X)$ (the Markov blanket of $X$). If there is no edge between $X$ and $Y$, it follows that $Y$ is the parent of a child of $X$ and is not a parent of $X$. Using $\Pi(X)$ to denote the parents of $X$, it therefore follows that $Y \perp X | \Pi(X)$, contradicting the assumption that there is no set $S$ such that $X \perp Y | S$. 

It follows that if there is no set $S$ such that $X \perp Y | S$, then $Y$ is either a parent or a child of $X$. In both cases, the edge $\langle X , Y \rangle$ is present in the skeleton of a faithful graph. \qed 

\begin{Th}\label{thveeimm}
Let ${\cal G}$ be a DAG that is faithful to a probability distribution $p$.  A vee structure $X-Y-Z$ (definition~\ref{defveestr}) is an immorality (definition~\ref{defimm}) if and only if there exists a subset $S \subset V\backslash \{X,Y,Z\}$ such that $X  \perp Z | S$. 

If $X-Y-Z$ is an immorality, then for any subset $S$ such that $X \perp Y|S$, it follows that  $X \not \perp Z | S \cup \{Y\}$. 
\end{Th}

\noindent {\bf Proof}  If there is no edge between $X$ and $Z$, it follows that $X \perp Z | S$ for some $S \subseteq V \backslash \{X,Z\}$. This is seen as follows. If $Z$ is not in the Markov blanket of $X$, then let $S = {\cal M}(X)$ (definition~\ref{defMB}). If $Z \in {\cal M}(X)$ but there is no edge between $X$ and $Z$, then $Z$ is the parent of a child of $X$. Let $S = \Pi(X) \cup \Pi(Z)$ (the parents of $X$ and the parents of $Z$), then $X \perp Z \|_{\cal G} S$ (that is, $X$ and $Z$ are $d$-separated given $S$), because any open trail from $X$ to $Z$ cannot take its first step through $\Pi(X)$ (instantiated chain or fork) and therefore takes its first step through a child of $X$ and (by symmetry) its last step through a child of $Z$. Therefore, all connections until the first instantiated node are chains (otherwise there is an uninstantiated collider) and hence the first instantiated node encountered is not in $\Pi(X)$ (otherwise the DAG contains a cycle). Symmetrically, the last collider encountered is not in $\Pi(Z)$. It follows that either the trail does not encounter any nodes in $\Pi(X) \cup \Pi(Z)$, but this is not possible, because the sequence of chain connections forces the last node on the trail to be a parent of $\Pi(Z)$ which is a contradiction. Otherwise the last collider node encountered is a parent of $X$, forcing $X$ to be a descendant of $Z$ and the first collider node (to prevent a cycle in the DAG) is a parent of $Z$, forcing $Z$ to be a descendant of $Z$, which is a contradiction. 

If $X-Y-Z$ is an immorality, then $Y \not \in \Pi(X) \cup \Pi(Z)$, for if it is in either of these sets, then it is both a parent and a child of at least one of the variables, which is a contradiction. If $S = \Pi(X) \cup \Pi(Z)$, then $X \perp Z \|_{\cal G} S$,   so there is a set $S$ satisfying the property. It is also clear that   $X \not \perp Z \|_{\cal G} S \cup \{Y\}$.

Now suppose that there exists a set $S \in V \backslash \{X,Y,Z\}$ such that $X \perp Z | S$. Since $Y \not \in S$, the trail $X-Y-Z$ with $Y$ uninstantiated is blocked and hence $X-Y-Z$ is a collider connection. 

If $X-Y-Z$ is an immorality, then it is clear that for any subset $S \in V \backslash \{X,Y,Z\}$ such that $X \perp Z | S$, $X \not \perp Z | S \cup \{Y\}$, because when $Y$ is instantiated, the trail $X-Y-Z$ is open when $X-Y-Z$ is a collider. \qed

\begin{Defn}[Markov Equivalence]\label{defmeq}
 Two directed acyclic graphs ${\cal G}_1 = (V, D_1)$ and ${\cal G}_2 = (V, D_2)$ are said to be Markov equivalent if and only if for any three disjoint subsets $A,B,S$ of $V$,
\[ A \perp B \|_{{\cal G}_1} S \Leftrightarrow A \perp B \|_{{\cal G}_2} S.\]
That is, $d$-separation statements in one graph are the same as $d$-separation statements in the other.
\end{Defn}

\begin{Th} \label{thmeq} Two directed acyclic graphs are Markov equivalent if and only if they have the same skeleton and the same immoralities.
\end{Th}

 \noindent {\bf Proof} The proof of this is found in P.Verma and J.Pearl, corollary 3.2 in~\cite{PV3}. \qed 

 \begin{Defn}[Essential Graph]\label{defesgr} 
Let ${\cal G}$ be a directed acyclic graph. The essential graph ${\cal G}^*$ associated with ${\cal G}$ is the graph with the same skeleton as ${\cal G}$, but where an edge is directed in ${\cal G}^*$ if and only if it occurs with the same orientation in every directed acyclic graph that is Markov equivalent to ${\cal G}$. 
\end{Defn}

\begin{Defn}[Strongly Protected Edge] \label{defstpred} An edge is said to be {\em strongly protected} if occurs either in an immorality, or else in one of the three structures found in figure~\ref{strongprotect}.
\end{Defn}

\begin{Lmm}\label{stprjust} Once the immoralities have been determined, the direction of the edges between $X$ and $Y$ in the structures of  figure~\ref{strongprotect} are determined. 
 \end{Lmm}

\noindent {\bf Proof of lemma~\ref{stprjust}} In the first structure, it is clear that if the edge $Z \mapsto X$ is directed from $Z$ to $X$, then the direction $X$ to $Y$ is forced, because otherwise there would be an additional immorality. 

In the second structure, it is clear that once the directions $X \mapsto Z$ and $Z \mapsto Y$ are forced, then the direction $X \mapsto Y$ is forced, because otherwise there is a cycle. 

In the third structure, if $(Z_1, Y, Z_2)$ is an immorality, then the direction $X \mapsto Y$ is forced, because if the edge is in the other direction, then either a new immorality $(Z_1, X , Z_2)$ is created, contradicting the conditional independence statements that have already been established, or else there is necessarily a cycle.

It is also clear that the structures in the definition~\ref{defstpred} are the only ones where the directions of edges are forced. The justification is as follows:  

Consider two neighbour nodes $X$ and $Y$ which are neighbours. Firstly, suppose that $X$ and $Y$ do not have any other common neighbours.  If $X$ does not have a neighbour $Z$ such that there is a directed edge $(Z , X)$, then the direction $(X , Y)$ is not forced; no additional immorality or cycle is created by either direction.

Now suppose that $X$ and $Y$ have at least one common neighbour. Suppose that there are no neighbours $Z$ such that both $(X, Z)$, $(Z , Y)$ in the directed edge set, then it is not necessary to force the direction $(X , Y)$ to prevent a cycle.

Suppose, furthermore, that there is no pair of common neighbours $W$ and $Z$, such that $(Z, Y)$ and $(W,Y)$ are both in the directed edge set and $(Z, X, W)$ is not an immorality. Then it is not necessary to force the direction $(X, Y)$ to prevent either   $(Z , X, W)$ becoming an immorality or else a cycle appearing. \qed

\setlength{\baselineskip}{2ex}

\end{document}